\documentclass[twocolumn]{aastex7}
\usepackage{txfonts}
\usepackage{graphicx}
\usepackage{subfigure}
\usepackage{lipsum}
\usepackage{stfloats}
\usepackage{float}
\usepackage{CJKutf8}
\usepackage{multirow}
\shorttitle{CCC of G178}
\shortauthors{Zhang et al.}

\begin{document}
\title{An S-shaped filament formed due to Cloud-Cloud Collision in molecular cloud G178.28-00.61}

\correspondingauthor{Tianwei Zhang \begin{CJK*}{UTF8}{gbsn}(张天惟)\end{CJK*} \email{twzhang@zhejianglab.org}} 

\author[0000-0002-1466-3484]{Tianwei Zhang}
\affiliation{Research Center for Astronomical computing, Zhejiang Laboratory, Hangzhou, China}
\email{twzhang@zhejianglab.org}

\author[0000-0002-5286-2564]{Tie Liu}
\affiliation{Shanghai Astronomical Observatory, Chinese Academy of Sciences, 80 Nandan Road, Shanghai 200030, People’s Republic of China}
\email{liutie@shao.ac.cn}

\author[0000-0002-5076-7520]{Yuefang Wu}
\affiliation{Department of Astronomy, Peking University, 100871, Beijing China}
\email{ywu@pku.edu.cn}

\author[0009-0003-4821-5502]{Linjing Feng}
\affiliation{National Astronomical Observatories, Chinese Academy of Sciences, Beĳing 100101, China}
\affiliation{University of Chinese Academy of Sciences, Beĳing 100049, China}
\email{ljfeng@nao.cas.cn}

\author[0000-0002-9151-1388]{Sihan Jiao}
\affiliation{National Astronomical Observatories, Chinese Academy of Sciences, Beĳing 100101, China}
\affiliation{Max Planck Institute for Astronomy, Konigstuhl 17, D-69117 Heidelberg, Germany}
\email{sihanjiao@nao.cas.cn}

\author[0000-0003-1140-2761]{Derek Ward-Thompson}
\affiliation{Jeremiah Horrocks Institute, University of Central Lancashire, Preston PR1 2HE, UK}
\email{DWard-Thompson@uclan.ac.uk}

\author[0000-0003-1665-6402]{Alessio Traficante}
\affiliation{INAF - Istituto di Astrofisica e Planetologia Spaziali (IAPS), Via Fosso del Cavaliere 100, I-00133 Roma, Italy}
\email{alessio.traficante@gmail.com}

\author[0000-0003-0972-1595]{Helen J Fraser}
\affiliation{School of Physical Sciences the Open University Milton keynes, MK7 6AA UK}
\email{helen.fraser@open.ac.uk}

\author[0000-0002-9289-2450]{James Di Francesco}
\affiliation{NRC Herzberg Astronomy and Astrophysics, 5071 West Saanich Road, Victoria, BC V9E 2E7, Canada}
\affiliation{Department of Physics and Astronomy, University of Victoria, Victoria, BC V8P 5C2, Canada}
\email{James.DiFrancesco@nrc-cnrc.gc.ca}

\author[0000-0002-6773-459X]{Doug Johnstone}
\affiliation{NRC Herzberg Astronomy and Astrophysics, 5071 West Saanich Road, Victoria, BC V9E 2E7, Canada}
\affiliation{Department of Physics and Astronomy, University of Victoria, Victoria, BC V8P 5C2, Canada}
\email{Douglas.Johnstone@nrc-cnrc.gc.ca}

\author[0000-0002-6622-8396]{Paul F. Goldsmith}
\affiliation{Jet Propulsion Laboratory, California Institute of Technology, 4800 Oak Grove Drive, Pasadena CA 91109, USA}
\email{paul.f.goldsmith@jpl.nasa.gov}

\author[0000-0001-8746-6548]{Yasuo Doi}
\affiliation{Department of Earth Science and Astronomy, Graduate School of Arts and Sciences, The University of Tokyo, 3-8-1 Komaba, Meguro, Tokyo 153-8902, Japan}
\email{doi@ea.c.u-tokyo.ac.jp}

\author[0000-0001-8315-4248]{Xunchuan Liu}
\affiliation{Shanghai Astronomical Observatory, Chinese Academy of Sciences, 80 Nandan Road, Shanghai 200030, People’s Republic of China}
\affiliation{Leiden Observatory, Leiden University, PO Box 9513, 2300, RA, Leiden, The Netherlands}
\email{liuxunchuan@shao.ac.cn}

\author[0000-0002-3179-6334]{Chang Won Lee}
\affiliation{Korea Astronomy and Space Science Institute 776, Daedeokdae-ro, Yuseong-gu, Daejeon, Republic of Korea 305-348}
\email{cwl@kasi.re.kr}

\author[0000-0001-5950-1932]{Fengwei Xu}
\affiliation{Kavli Institute for Astronomy and Astrophysics, Peking University, Beijing 100871, People's Republic of China}
\affiliation{Department of Astronomy, Peking University, 100871, Beijing China}
\email{fengwei.astro@pku.edu.cn}

\author[0000-0002-6740-7425]{Ram K. Yadav}
\affiliation{National Astronomical Research Institute of Thailand (NARIT), Sirindhorn AstroPark, 260 Moo 4, T. Donkaew, A. Maerim, Chiangmai 50180, Thailand}
\email{ramkeshyadav2005@gmail.com}

\author[0000-0002-7126-691X]{Glenn J White}
\affiliation{Department of Physics \& Astronomy, The Open University, Milton Keynes MK7 6AA, England}
\affiliation{RAL Space, The Rutherford Appleton Laboratory, Chilton, Didcot, Oxfordshire OX11 9DL, England}
\email{gjw255@open.ac.uk}

\author[0000-0002-9574-8454]{Leonardo Bronfman}
\affiliation{Departamento de Astronomía, Universidad de Chile, Casilla 36-D, Santiago, Chile}
\email{leo@das.uchile.cl}

\author[0000-0002-4336-0730]{Yi-Jehng Kuan}
\affiliation{Department of Earth Science, National Taiwan Normal University, Taipei, 11677, Taiwan, ROC}
\email{kuan@ntnu.edu.tw}

\author[0000-0003-2412-7092]{Kee-Tae Kim}
\affiliation{Korea Astronomy and Space Science Institute 776, Daedeokdae-ro, Yuseong-gu, Daejeon, Republic of Korea 305-348}
\email{ktkim@kasi.re.kr}

\author[0000-0003-4811-2581]{Donghui Quan}
\affiliation{Research Center for Astronomical computing, Zhejiang Laboratory, Hangzhou, China}
\email{donghui.quan@zhejianglab.org}

\begin{abstract}
We present compelling observational evidence supporting G178.28-00.61 as an early-stage candidate for Cloud-Cloud Collision (CCC), with indications of the formation of an S-shaped filament, evenly-separated dense cores, and young star clusters. The observations of CO molecular line emission demonstrate the existence of two interacting molecular clouds with systemic velocities of 0.8 km~s$^{-1}$ and -1.2 km~s$^{-1}$, respectively. The convergence zone of these two clouds reveals an S-shaped filament in the JCMT 850 $\mu$m continuum image, suggesting cloud interaction. In line with expectations from CCC simulations, broad bridging features are discernible in the position-velocity diagrams. An elevated concentration of identified Class I and II young stellar objects along the filament at the intersection area further supports the hypothesis of a collision-induced origin.
This observation could be explained by a recent MHD model of CCC (Kong et al. 2024), which predicts a similar morphology, scale, density, and unbound status, as well as the orientation of the polarization.

\end{abstract}


\section{Introduction}
\label{sec:intro}
Cloud-cloud collisions (hereafter CCC) represent a vigorous interaction process between dense molecular clouds within the interstellar medium (ISM), which deviates from traditional star formation theories, e.g., \cite{vaz19, pad20}. These collisions, however, have the potential to facilitate filament formation and enhance star formation \citep{haw15, wu17}.

CCC might shed light on the filament formation mechanism, which is pivotal for comprehending the very early stage of star formation. Investigations of nearby clouds conducted by Herschel have revealed the ubiquity of filamentary structures within the cold ISM \citep{and14} and the Galactic Plane \citep{sch20}. The presence of a primary filament, accompanied by an array of orthogonal striations, appears to be a prevalent configuration among molecular clouds. Filaments have been effectively predicted in simulations of supersonic turbulence in the absence of gravity, which can produce hierarchical structures with an observed lognormal density distribution \citep{vaz94}. The formation of filaments through self-gravitational fragmentation of sheet-like clouds has also been observed in simulations of one-dimensional compression \citep{vaz19}, such as that induced by an expanding HII region, an old supernova remnant, or the collision of two clouds. Furthermore, more than 70\% of the cores are situated within the filaments \citep{and14}. In summary, filaments play a critical role in star formation \citep{hac23}. Nevertheless, the mechanisms by which filaments form within the cold ISM remain inadequately understood. One hypothesis proposes that filament formation may be induced by CCC at the dense colliding interface \citep{fuk21}.

At the same time, CCC may create exceptional conditions that enhance the rate of star formation \citep{fuk21}. As noted in \cite{tra20}, attaining a critical value of the surface density ($\Sigma >$0.1 g~cm$^{-2}$ at the pc scales) is essential to counteract turbulence and sustain global gravitational collapse, thereby facilitating the mass accretion and star formation.  Such high density could be a consequence of the CCC and subsequently trigger low-mass and high-mass star formation.

Recently, several simulations aim to elucidate the characteristics of CCC. \cite{hen08} claimed that the molecular clouds are formed by moderate supersonic collision of warm atomic gas streams. \cite{wu17} examined the interaction between two identical clouds (n$_H$=100~cm$^{-3}$, R=20~pc) with varying relative velocities of each other within a range of 20 km~s$^{-1}$. The study employed the three-dimensional magnetohydrodynamics numerical code Enzo, with a timescale of 4~Myr commencing from the initial contact of the two clouds. The findings indicate that, after a few million years, filamentary structures are dominant in all scenarios. Unsurprisingly, the colliding scenarios distinctly develop a greater number of cores and higher densities compared with the scenario without collisions (v=0 case). In instances where collisions occur, higher relative velocities result in the formation of stronger shocks, leading to regions of higher-density and the formation of more massive clumps at earlier stages. Another simulation \citep{tak14} investigated the collision between distinct clouds, specifically a large cloud (n$_H$=25.3~cm$^{-3}$, R=7.2~pc) and a small cloud (n$_H$=47.4~cm$^{-3}$, R=3.5~pc), resulting in the creation of a compressed layer which subsequently facilitates the formation of high-mass stars.
\cite{kon24} have simulated the collision of two identical atomic gas clouds via the mechanism of collision-induced magnetic reconnection (CMR), predicting the formation of molecular clouds and filaments that are supersonic turbulence dominated and gravitationally unbounded. Despite the adoption of markedly different initial conditions, the enhancement of filament and star formation due to CCC is evident across various simulations.

Nonetheless, observationally, the verification of a collision event presents considerable challenges due to the short timescale of collisions (generally a few Myr), the infrequent occurrence of such events in the Milky Way (approximately once every 10 Myr), and the disruption of progenitor clouds \citep{haw15}. Consequently, any prospective candidate for CCC is of significant value and merits thorough investigation.

Observational evidence of CCC has been documented for over four decades. Several well-known star-forming regions, which exhibit two distinct velocity components attributable to two converging clouds, have been identified as potential instances of CCC. Such phenomena were initially suggested in LkH$\alpha$~198, which is associated with Herbig Ae/Be stars \citep{lor77}. Additionally, \cite{dic78} identified that DR21/W75, a region of high-mass star formation, was generating an OB2 association at the CCC interface, concomitant with the production of an HII region. As reported by \cite{ode92}, GR110-13 was in the process of forming two B-type stars as a result of CCC. Moreover, the collision in W49 resulted in the formation of a star cluster and an HII region \citep{buc96}. Recently, a number of complex sources have been investigated and are anticipated to demonstrate evidence of CCC. For instance, the W2 region exhibits two molecular lobes \citep{fur09, oha10} undergoing collision at a significant relative speed of 20 km~s$^{-1}$, which have subsequently initiated the formation of a massive star cluster. The famous Galactic HII region, the Trifid Nebula (M20), displays two clouds converging toward the central O-type star with a relative velocity of 10 km~s$^{-1}$, providing strong evidence for CCC induced star formation \citep{tor11, haw15}. Moreover, \cite{tor15} claimed that the RCW120, resembling an open bubble, might have resulted from CCC, as evidenced by the formation of OB stars along the bubble's edge and an HII region within it. These prior investigations share the characteristic that their sources are associated with stars or star clusters. Relatively few early-stage clouds induced by CCC, however, have been subjected to detailed study, especially those have not yet shown HII regions \citep[e.g.,][]{nak14,gon17,iss20}. Observational samples of CCC-induced low- to intermediate-mass star formation remain scarce. Our newly proposed CCC candidate, G178.28-00.61 (hereafter G178), identified as a Planck Cold Clump at a distance of 0.94~kpc, remains in an early evolutionary stage and not yet affected by stellar feedback \citep{wu12,zha16}, rendering it particularly significant for elucidating the initial conditions of CCC.

In this paper, we present new continuum emission and molecular line observations toward G178. The observations are detailed in Section \ref{sec:observation}. Section \ref{sec:result} describes the primary results of the continuum data and spectral analysis, and Section \ref{sec:discussion} explores the evolutionary states, evidence for CCC, and unresolved challenges. In Section \ref{sec:summary}, we outline our conclusions.

\section{Observations}
\label{sec:observation}
\subsection{Continuum Data}
The continuum data of G178 were obtained in 2016 as a part of the SCUBA-2 Continuum Observations of Pre-protostellar Evolution (SCOPE) project \citep{liu18}, a legacy survey conducted by the James Clerk Maxwell Telescope (JCMT). Within this survey, continuum emission at wavelengths of 850 $\mu$m and 450 $\mu$m was concurrently surveyed using the constant-velocity (CV) Daisy mapping mode by the Submillimetre Common-User Bolometer Array \citep[SCUBA-2;][]{hol13}. The instrument has a field of view of 45 arcmin$^2$. The FWHM (full width at half-maximum) main beam size of SCUBA-2 is 7$\arcsec$.9 at 450 $\mu$m and 14$\arcsec$.1 at 850 $\mu$m. With an exposure time of approximately 16 minutes per map, the rms noise level in the most sensitive central 3$\arcmin$ area of our target attained approximately 8~mJy per pixel (1~pixel=4$\arcsec$). The acquired SCUBA-2 data were calibrated and processed using the SMURF software within the Starlink package \citep{cha13}.

In addition, the Herschel, Spitzer, and WISE Space Telescopes previously targeted this object, with data available from the NASA/IPAC Infrared Science Archive (IRSA).

The Herschel data were acquired from the Herschel Science Archive\footnote{\href{http://archives.esac.esa.int/hsa/whsa/}{http://archives.esac.esa.int/hsa/whsa/}}. The expansive region encircling our target was systematically mapped using both PACS at 70 $\mu$m and 160 $\mu$m, and SPIRE at 250 $\mu$m, 350 $\mu$m, and 500 $\mu$m, as part of the Hi-GAL survey \citep{mol10}. All datasets have undergone deep processing, reaching level 2.5 for PACS and level 3 for SPIRE.

The WISE archive\footnote{\href{https://irsa.ipac.caltech.edu/Missions/wise.html}{https://irsa.ipac.caltech.edu/Missions/wise.html}} images encompass four spectral bands at wavelengths of 3.4 $\mu$m, 4.6 $\mu$m, 12 $\mu$m, and 22 $\mu$m, corresponding to angular resolutions of 6.1$''$, 6.4$''$, 6.5$''$, and 12$''$, respectively. Furthermore, the AllWISE Catalogue available in the VizieR Online Data Catalog \citep{cut13} was examined for single point sources, yielding evidence supporting the presence of young stellar objects (YSOs) in G178.

\subsection{Spectral Line Observations}
Observations of the $^{12}$CO, $^{13}$CO, and C$^{18}$O J=1-0 spectral lines were conducted utilizing the 13.7~m telescope at Purple Mountain Observatory (PMO) in December, 2011. The principal findings have been documented in \cite{wu12} and \cite{zha16}. The beam size is around 50$\arcsec$ in the 115~GHz frequency band. The velocity resolution of the $^{12}$CO~(1-0) line is 0.17~km~s$^{-1}$, while that of $^{13}$CO~(1-0) and C$^{18}$O~(1-0) is 0.16~km~s$^{-1}$. The On-the-fly (OTF) observing mode was employed to create a 22$\arcmin\times22\arcmin$ map. The typical rms was 0.5~K in T$^{*}_{A}$ for $^{12}$CO~(1-0), and 0.3~K for $^{13}$CO~(1-0) and C$^{18}$O~(1-0). 

Subsequently, the $^{12}$CO, $^{13}$CO, and C$^{18}$O J=2-1 lines were observed utilizing the Caltech Submillimeter Observatory (CSO) in January, 2014. The main beam efficiency was recorded as 76\%. The half-power beam width (HPBW) was measured at 32$\arcsec\pm3\arcsec$ within the 230~GHz frequency band. The velocity resolution for the $^{12}$CO~(2-1) line is 0.35~km~s$^{-1}$, whereas the resolution for the $^{13}$CO~(2-1) and C$^{18}$O~(2-1) lines is 0.37~km~s$^{-1}$. The OTF mapping covered a size of 5$\arcmin\times5\arcmin$. The typical root mean square (rms) noise levels were determined to be 0.35~K, 0.31~K, and 0.30~K in T$^{*}_{A}$ for $^{12}$CO~(2-1), $^{13}$CO~(2-1), and C$^{18}$O~(2-1), respectively. The analysis was conducted using the software package GILDAS/CLASS \citep{gui00}.

\begin{figure*}
    \begin{minipage}[c]{\textwidth}
	\centering
	\includegraphics[width=\textwidth]{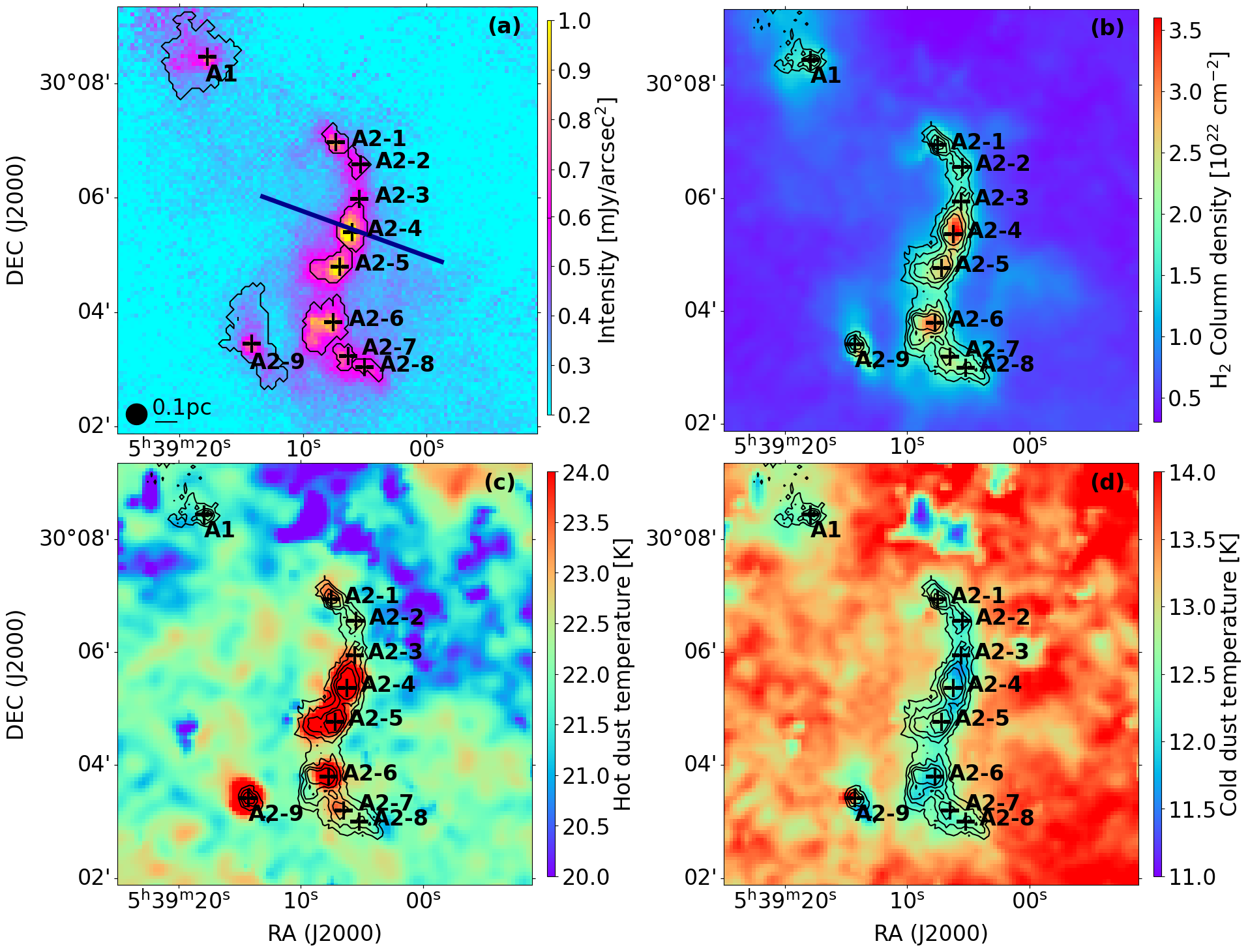}
    \end{minipage}
    \centering
    \begin{minipage}[c]{0.9\textwidth}
	\includegraphics[width=\textwidth]{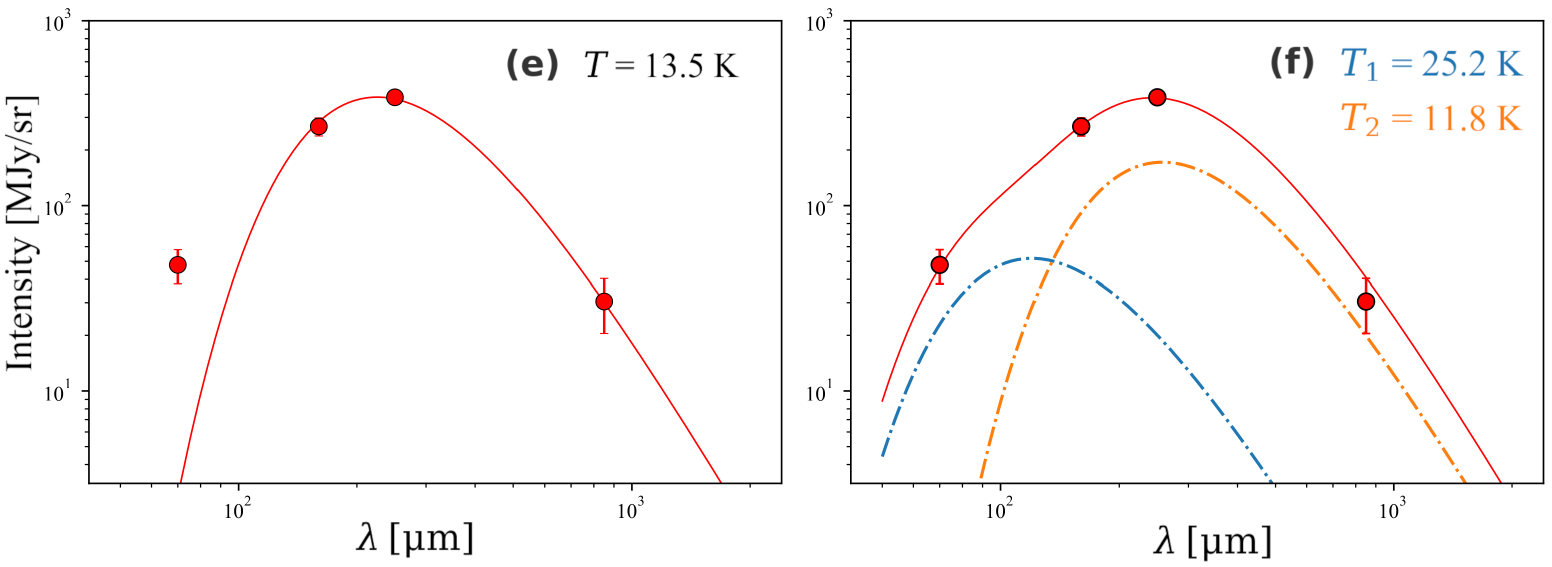}
    \end{minipage}
    \caption{(a) The 850 $\mu$m continuum emission map of G178.2-0.6 in color scale, detected by the JCMT and Planck. Ten emission peaks are marked, which were subject to subsequent spectral observation using the KVN in single point mode. The cores identified with the dendrogram are delineated by black contours. The JCMT beam size and the corresponding physical scale are shown in the bottom left corner. The dust polarization is shown by the dark blue solid line at the center, roughly perpendicular to the filament. (b) to (d) The pixel-by-pixel SED fitting results for two-component modified grey body spectra, based on continuum images across four bands (70, 160, 250 $\mu$m Herschel data and 850 $\mu$m JCMT+Planck data). (b) The total column density map, overlaid with 850 $\mu$m continuum contours which range over 0.47--0.77 mJy/arcsec$^2$ with a step of 0.1 mJy/arcsec$^2$. (c) and (d): the derived dust temperature of the warm and cold component, respectively. (e) and (f) The example SEDs of core A2-4 with single-component and two-component fit.}
    \label{fig:continuum-SED}
\end{figure*}

A follow-up study involving simultaneous multifrequency observations using the Korean VLBI Network (KVN) \citep{liu18} contributes to an enhanced comprehension of G178. Utilizing multifrequency band receiver systems at 22 GHz, 43 GHz, 86 GHz, and 129 GHz, we conducted simultaneous observations of five molecular lines in a single pointing mode using single-dish performance of KVN in March, 2016. The ten observed points were chosen from the 850 $\mu$m continuum map as the peak positions, depicted as black crosses in Figure \ref{fig:continuum-SED}. Spectral observations specifically covered lines of H$_2$O (6-5), CH$_3$OH (7-6), HCO$^+$ (1-0), N$_2$H$^+$ (1-0), and H$_2$CO ($2_{1,2}-1_{1,1}$), with respective velocity resolutions of 0.21 km s$^{-1}$, 0.11 km s$^{-1}$, 0.05 km s$^{-1}$, 0.05 km s$^{-1}$, and 0.03 km s$^{-1}$. The spatial resolutions range from 28" to 2.7'. The system temperatures varied between 116 K and 231 K. The main beam efficiency was approximately 50\%.

\section{Results}
\label{sec:result}
\subsection{Continuum emission}
\subsubsection{850 $\mu$m continuum and core identification}
In order to complement the missing flux in the JCMT observation, we adhered to the methodology outlined in \cite{jia22} to amalgamate the JCMT 850 $\mu$m image with the 850 $\mu$m image procured by the Planck Space Observatory, employing this combined JCMT+Planck 850 $\mu$m image for the following discussion. The combined 850 $\mu$m image is shown in Figure \ref{fig:continuum-SED} (a), revealing a well-defined S-shaped filamentary structure. Within the observed region, ten emission peaks have been identified along the filamentary structure (labeled in Figure \ref{fig:continuum-SED} and coordinates noted in Table \ref{table:SED}), which were utilized for subsequent KVN single-point observations. \cite{zha16} identified two CO clumps in G178. The northeastern clump, which appears spatially isolated from other components, is designated as A1, whereas the remaining components (A2-1 to A2-9) are assigned as part of the A2 clump. The emission peaks within the A2 clump exhibit a pattern of regular spacing, which serves as a typical representation of sausage fragmentation (refer to Section \ref{sec:fragmentation} for a more detailed discussion). Along the filament, the emission from the central pixels (i.e., A2-4, A2-5) is approximately one order of magnitude stronger than those located on either side of the filament, suggesting the presence of a possible mass gradient and evolutionary gradient discussed in Section \ref{sec:evolution}.

We have used the dendrogram\footnote{\href{https://dendrograms.readthedocs.io/en/stable/}{https://dendrograms.readthedocs.io/en/stable/}} algorithm to identify and build a catalogue of compact sources. For G178, the detection threshold is set to 3$\sigma$; the significance value is 1$\sigma$; and the minimum source size is adopted to be 15 pixels (half of the beam size). $\sigma$ is the rms noise level (0.0114~mJy/arcsec$^2$), which is calculated based on the sigma-clipped standard deviation of pixel values in the whole image. The result gives 9 determined cores (shown as black contours in Figure \ref{fig:continuum-SED} (a)). All the previously noted emission peaks are associated with a dendrogram cores, except peak A2-3. The effective core radii given in the fourth column of Table \ref{table:SED} are calculated by $\sqrt{(A/\pi)}$, where A is the defined polygon area of the core. The core radii (9.30$''$--33.55$''$) correspond to 0.04--0.15~pc, close to the typical core scale of 0.1~pc. Subsequently, the integrated 850 $\mu$m fluxes within area A of each core are summarized in the 5th column of Table \ref{table:SED}, ranging from 0.11 to 0.67 Jy, peaking at core A2-6.

\startlongtable
\begin{deluxetable*}{cccccccccccc}
\setlength{\tabcolsep}{0.08in}
\tabletypesize{\scriptsize}
\tablewidth{0pt}
\tablecaption{Parameters of ten 850~$\mu$m continuum peaks and corresponding cores \label{table:SED}}
\tablehead{
Name & RA(J2000) & DEC(J2000) & Radius & Flux & N$_{H_{2}}$ & n$_{H_{2}}$ & $\Sigma$ & $T_{SED}$ & M$_{SED}$ & M$_{jeans}$ & Associated stellar objects\\
     &(h m s)   &(d m s) & $\arcsec$& Jy & 10$^{22}$cm$^{-2}$ & 10$^{4}$cm$^{-3}$ & g~cm$^{-2}$ & K & M$_\odot$ & M$_\odot$ & }
\startdata
A1   & 05 39 17.90 & 30 08 26.3& 33.55 &0.57 & 1.29(0.04) & 2.06(0.06) & 0.06 & 12.7(0.4) & 21.3(0.7) & 80.6(2.4) & -- \\
A2-1 & 05 39 07.50 & 30 06 56.3& 11.28 &0.20 & 1.74(0.03) & 8.21(0.16) & 0.08 & 12.6(0.3) & 3.2(0.1) & 5.5(0.1) & Class I YSO\\
A2-2 & 05 39 05.50 & 30 06 33.3& 9.30 &0.11  & 1.71(0.03) & 9.78(0.19) & 0.08 & 12.2(0.3) & 2.2(0.1) & 7.8(0.1) & Class I YSO\\
A2-3 & 05 39 05.60 & 30 05 56.7 &--  &--   & 1.98(0.03) & --         & -- & 12.0(0.3) & --         & --         & Class I YSO\\
A2-4 & 05 39 06.20 & 30 05 21.9& 17.19 &0.60 & 2.71(0.02) & 8.41(0.07) & 0.13 & 12.1(0.2) & 11.7(0.1) & 18.6(0.2) & two Class I YSOs\\
A2-5 & 05 39 07.20 & 30 04 45.9& 16.12 &0.47 & 2.30(0.03) & 7.61(0.09) & 0.11 & 12.5(0.2) & 8.7(0.1) & 38.8(0.6) & Class I\&II YSO\\
A2-6 & 05 39 07.70 & 30 03 47.9& 22.68 &0.67 & 2.26(0.03) & 5.32(0.07) & 0.11 & 12.2(0.2) & 17.0(0.2) & 6.9(0.1) & Class I YSO\&shock\\
A2-7 & 05 39 06.50 & 30 03 11.9& 9.84  &0.12 & 2.19(0.03) & 11.89(0.17) & 0.10 & 12.4(0.2) & 3.1(0.1) & 7.5(0.1)& shock knot\\
A2-8 & 05 39 05.20 & 30 03 00.1& 9.57  &0.11 & 2.00(0.04) & 11.16(0.20) & 0.09 & 12.6(0.2) & 2.7(0.1) & 7.8(0.2) & -- \\
A2-9 & 05 39 14.30 & 30 03 24.9& 31.19 &0.51 & 1.25(0.03) & 2.13(0.06) & 0.06 & 12.8(0.4) & 17.7(0.5) & 4.2(0.2) & Class I YSO\\
\enddata
\end{deluxetable*}
\vspace{-10pt}

\subsubsection{SED fitting results}
In order to ascertain the column densities and dust temperatures, we perform two-component, modified blackbody spectral energy distribution (SED) fits for each pixel using data of four frequency band, assuming one high-temperature dust component resulting from protostellar heating and a low-temperature component that represents the cold filament in the background. Together with the combined JCMT+Planck 850 $\mu$m image, data from the Herschel PACS at 70 $\mu$m, 160 $\mu$m, and SPIRE at 250 $\mu$m are incorporated for the purpose of SED fitting, while discarding the 350 $\mu$m and 500 $\mu$m data to obtain high resolution maps. For each Herschel band, we first cropped the Hi-GAL-178 images to our field, and then gaussian convolved them to a common angular resolution of the largest telescope beam (18.2$\arcsec$ for 250 $\mu$m), and then regridded them to the same pixel size (6.0$\arcsec$).

Finally, we apply the modified grey body function for fitting:
\begin{equation}
S_{\nu}=\frac{2h\nu^3}{c^2(e^{\frac{h\nu}{kT}}-1)}(1-e^{-\tau_\nu}),
\label{eq_GB}
\end{equation}
where
\begin{equation}
\tau_\nu=\mu_{H_2}m_H \kappa_{\nu} N_{H_2},
\end{equation}
where $\mu_{H_2}$=2.8 is the mean molecular weight at cosmic abundances of X=0.71, Y=0.27, and Z=0.02 \citep{cox00,kau08}, $m_H$ is the mass of hydrogen, N$_{H_{2}}$ is the column density, and $\kappa_{\nu}$ is the dust opacity, which follows the power-law $\kappa_{\nu}=\kappa_0 (\frac{\nu}{\nu_0})^{\beta}$. 
We adopt $\kappa_0$ = 0.1 $\mathrm{cm^2g^{-1}}$ and $\beta$ = 1.8 at 250~$\mu$m \citep{1983QJRAS..24..267H} under the assumption that the gas-to-dust ratio equals 100. During the fitting process, we first tested single-component fit and found the data point at shortest wavelength 70 $\mu$m could not be well fitted (see Fig. \ref{fig:continuum-SED} (e) for example). Therefore, we added two grey body spectra (Equation \ref{eq_GB}) together to perform a two-component grey body spectrum fit, leading to reasonable results (see Fig. \ref{fig:continuum-SED} (f) for example). Subsequently, we obtained gas column density maps and dust temperature maps corresponding to cold (filament temperature $\sim$ 12 K) and warm (temperature $\sim$ 24 K) components of the grey body radiation.

For the G178 cloud, we obtained the total column density map and dust temperature, as depicted in Figure \ref{fig:continuum-SED} (b) to (d). The N$_{H_{2}}$ provides a summary of the two grey body components, wherein the hot component contributes just $\sim$1\% of the material. Therefore, the temperature of the cold component represents the physical condition of essentially all the dust and these values are adopted as T$_{dust}$. The column density distribution is generally consistent with the 850 micron continuum map, effectively capturing the majority of the dust emission. The temperature map represents the cold grey body component, demonstrating lower temperatures in the filament compared to the diffuse regions, with the minimum temperature recorded in core A2-4 (11.6 K). The values of parameters N$_{H_{2}}$ and T$_{dust}$ were averaged within the polygon region of each core and are documented in columns 6 and 9 of Table \ref{table:SED}, respectively. The H$_2$ column densities are calculated as 1.25--2.71 $\times$ 10$^{22}$ cm$^{-2}$ with a general error of 0.04 $\times$ 10$^{22}$ cm$^{-2}$, corresponding to surface density of 0.06--0.13 g~cm$^{-2}$, which peak at core A2-4. The temperatures of the cold component are 12.0--12.8~K, with an error of 1.1~K, peaking at core A2-9. The number densities n$_{H_2}$ are derived assuming the cores are spherical clouds, and range over 0.2--1.2 $\times$ 10$^{5}$ cm$^{-3}$.

The SED results provide basic physical information for subsequent evaluations. Specifically, the dust temperature is compared with gas temperature derived by CO emission. And the N$_{H_{2}}$ derived from dust serves as the standard reference to calculate molecular fractional abundance in the following sections.

\subsubsection{Masses}
\label{sec:mass}
The dust mass (M$_{SED}$) is determined as the aggregate of H$_2$ column densities within the core area, as derived from dust SED results, and those of the cores studied here are presented in Table \ref{table:SED}. The temperatures utilized in our study align closely with the standardized dust temperature of 13 K adopted in the previous research of Planck Cold Clumps \citep{ede19}. The resulting total masses derived from dust emission range over 2.2--21.3~M$_{\odot}$ with a mean value of 9.7$\pm$7.5 M$_{\odot}$. Following the methodology outlined in \cite{zha16}, Jeans masses including thermal and non-thermal motions are computed utilizing the kinematic temperatures and velocity dispersions obtained from C$^{18}$O (1-0). The computed Jeans masses are 4.2--80.6~M$_{\odot}$. Cores A2-6 and A2-9 exhibit M$_{SED}$ values exceeding twice M$_{jeans}$, indicating that these cores may be self-gravitationally bound and predisposed to further fragmentation. In contrast, the remaining seven cores have M$_{SED}$ values below M$_{jeans}$, with M$_{jeans}$/M$_{SED}$ ratios predominantly exceeding 2, suggesting that these cores may be gravitationally unbound and not going to fragment. This finding contradicts the observation of core A2-5, which, despite a M$_{jeans}$/M$_{SED}$ ratio of $\sim$4.4, is associated with two young stellar objects (see Section \ref{sec:YSO} for more discussion), implying the presence of at least two fragments. Higher resolution observation is required to investigate whether there is further fragmentation or not.

\begin{figure*}
    \begin{minipage}[c]{\columnwidth}
	\centering
	\includegraphics[width=\columnwidth]{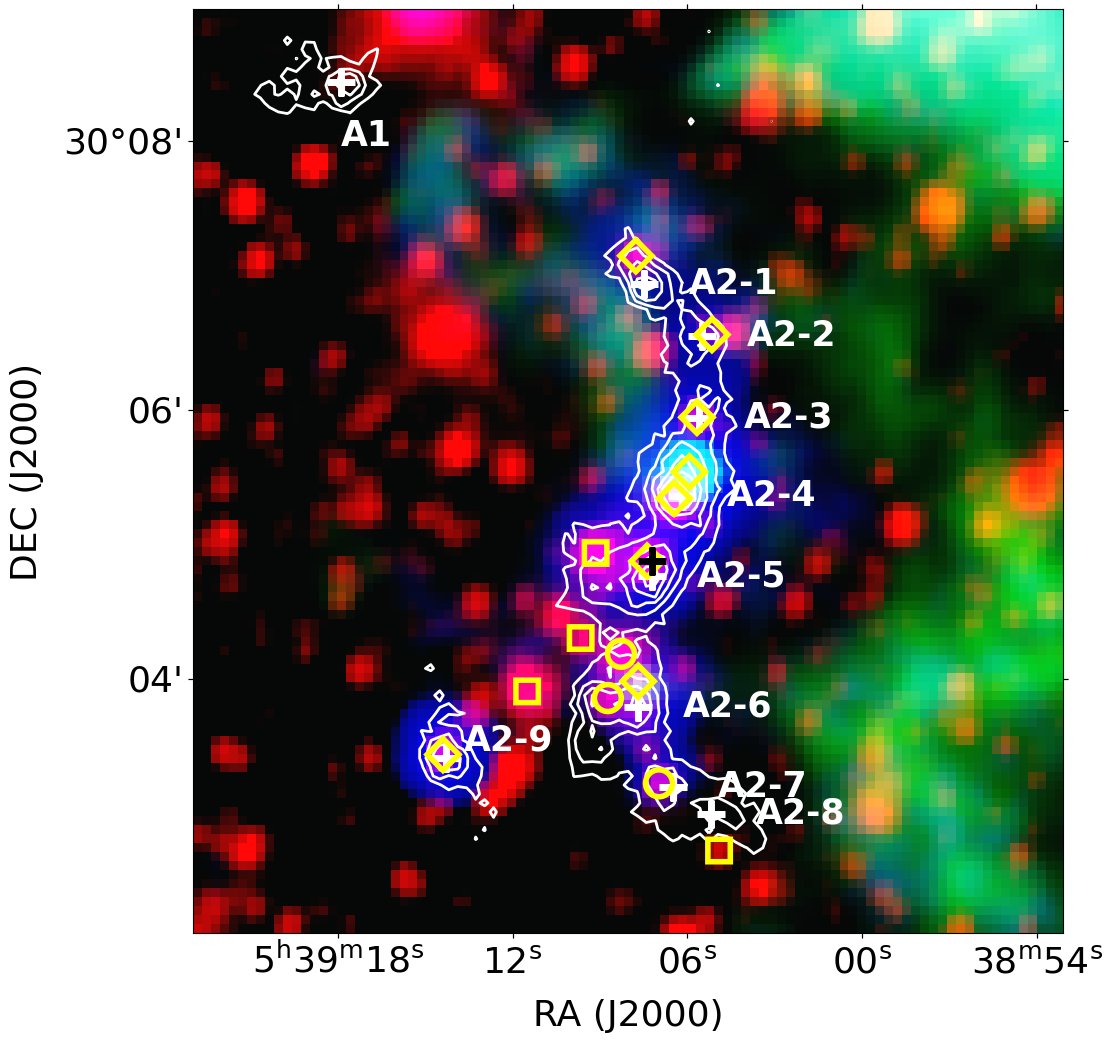}
    \end{minipage}
    \begin{minipage}[c]{\columnwidth}
        \centering
        \includegraphics[width=0.9\columnwidth]{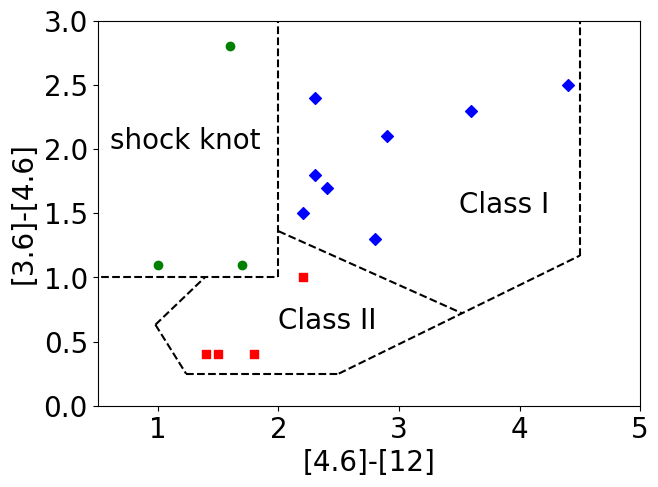}
    \end{minipage}
    \caption{Left panel: WISE three-color composite image (RGB: 22~$\mu$m, 12~$\mu$m, 3.4~$\mu$m), overlaid with associated YSOs and 850~$\mu$m continuum contours. Class I YSOs are depicted in yellow diamonds, Class II YSOs in squares, shock knots in circles. The black cross shows the IRAS point source. The continuum peaks are marked as white crosses and names on the side. Right panel: The color–color diagram for the bands 3.4~$\mu$m, 4.6~$\mu$m, and 12~$\mu$m used to classify YSOs associated with G178. The dashed lines denotes the criteria used by \cite{koe12,koe14} to delineate the YSO classes.}
    \label{fig:WISE_YSO}
\end{figure*}

\subsection{Associated stellar objects}
\label{sec:YSO}
The young stellar objects associated with G178 were analyzed utilizing infrared data. Figure \ref{fig:WISE_YSO} shows an RGB image of the G178 filament, which was generated using WISE W1 (3.4~$\mu$m) for blue, W3 (12~$\mu$m) for green, and W4 (22~$\mu$m) for red. The 12~$\mu$m map was modified to remove significant artifacts \citep{mei14}. Within the Simbad database, only one IRAS source was identified, which corresponds to Core A2-4, marked by a black cross in Figure \ref{fig:WISE_YSO}. Nonetheless, numerous other bright infrared sources associated with dense cores are visible in the WISE three-color image. Subsequently, the AllWISE catalog was consulted, and point sources within the field of view were classified and marked in Figure \ref{fig:WISE_YSO} based on their types. The criteria used for classifying YSOs were derived from \cite{koe14}, while shock knots resulting from outflows colliding with cold material were classified according to \cite{koe12}. Based on the color-color diagram (see the lower panel in Figure \ref{fig:WISE_YSO}), eight sources were classified as Class I YSOs, while four were designated as Class II YSOs. Additionally, three shock knots were identified. Most continuum peaks are associated with infrared sources, except for A1 and A2-8 (see the last column in Table \ref{table:SED}). For the cores located at the center of the filament, A2-4 and A2-5, each contains two close ($\sim$0.02~pc) and bright YSOs. Core A2-6 hosts one Class I YSO and two shock knots, giving the possibility of YSO-powering outflows. For the core A2-7, one associated shock knot is determined without YSO, hence it could be powered by another core, e.g., A2-6. There are three Class II YSOs not associated with any continuum peaks, suggesting a migration from the original cores possibly due to CCC.

We intended to find outflows to confirm the detection of YSOs. However, no HH objects appear within a radius of 10 arcmin, since G178 is a young and deeply embedded cloud. Besides, the $^{12}$CO spectra show multiple peaks and broad line widths (see example spectra in Figure \ref{fig:co-spec}), which may partially attribute to outflows. However, line wings could not be easily determined due to complex velocity components. In the future, we need observation of other outflow tracers, e.g., SiO, to confirm the existences of outflows and support the identification of YSOs.

\subsection{Molecular lines and Mapping results}
\subsubsection{CO J=1-0 and J=2-1}
The integrated intensity maps of $^{13}$CO (1-0) represented in color, and C$^{18}$O (1-0) illustrated as contours, are depicted in the Figure \ref{fig:CO} (a), both highlighting the densest regions of the cloud. Similarly, the Figure \ref{fig:CO} (b) depicts analogous mappings for the J=2-1 transitions, which are indicative of cores exhibiting even greater density. Nonetheless, the CSO observational field of view is restricted, consequently excluding core A1. The 850 $\mu$m continuum shows a significant association with the most intense region of CO emission, though there are some CO cores showing no corresponding continuum cores. For example, the continuum peak A2-9 is not located in the main CO filament, which may indicate CO depletion at this position.

\begin{figure*}%[h]
	\begin{minipage}[c]{0.24\linewidth}
		\centering
		\includegraphics[width=\textwidth]{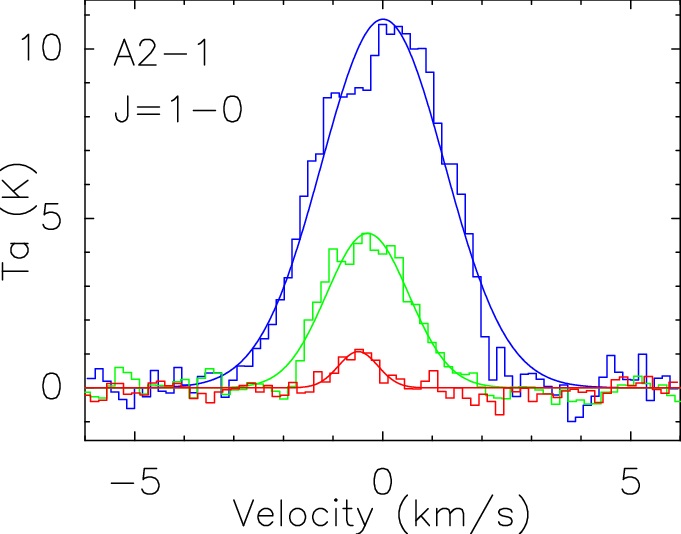}
	\end{minipage}
	\begin{minipage}[c]{0.24\linewidth}
		\centering
		\includegraphics[width=\textwidth]{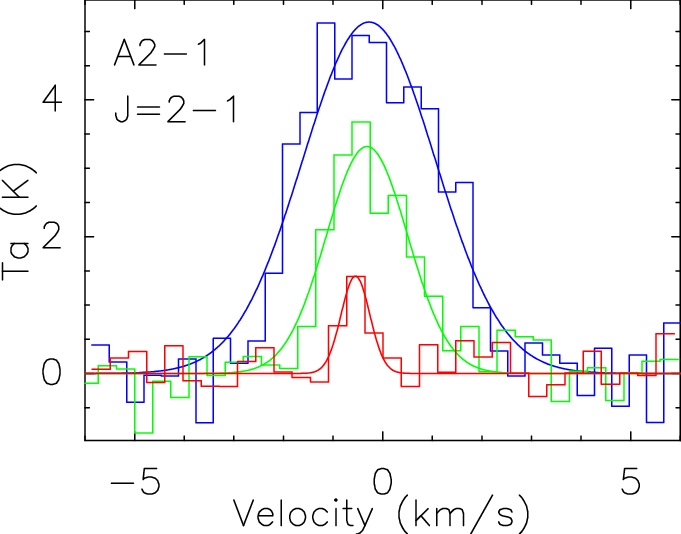}
	\end{minipage}
    	\begin{minipage}[c]{0.24\linewidth}
		\centering
		\includegraphics[width=\textwidth]{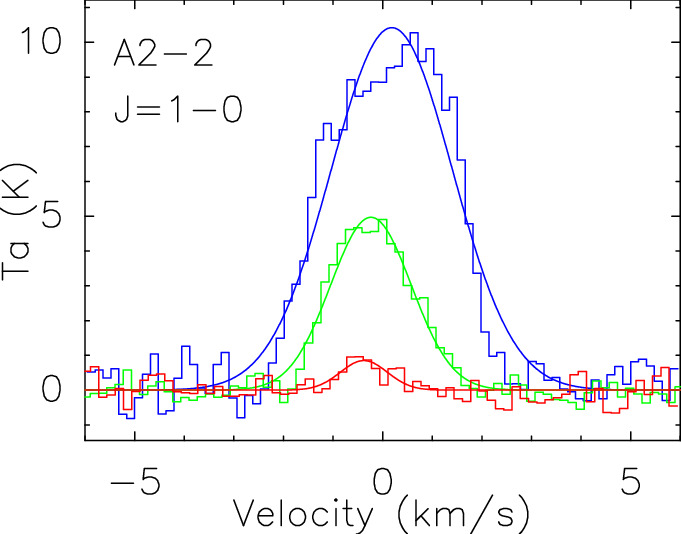}
	\end{minipage}
	\begin{minipage}[c]{0.24\linewidth}
		\centering
		\includegraphics[width=\textwidth]{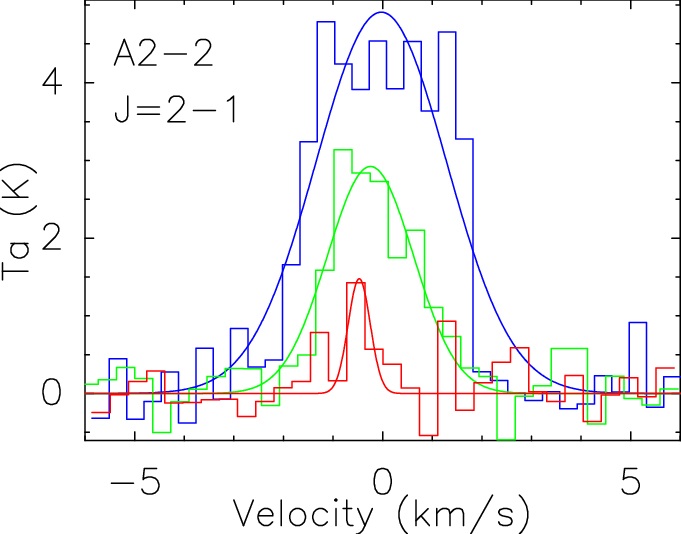}
	\end{minipage}
    	\begin{minipage}[c]{0.24\linewidth}
		\centering
		\includegraphics[width=\textwidth]{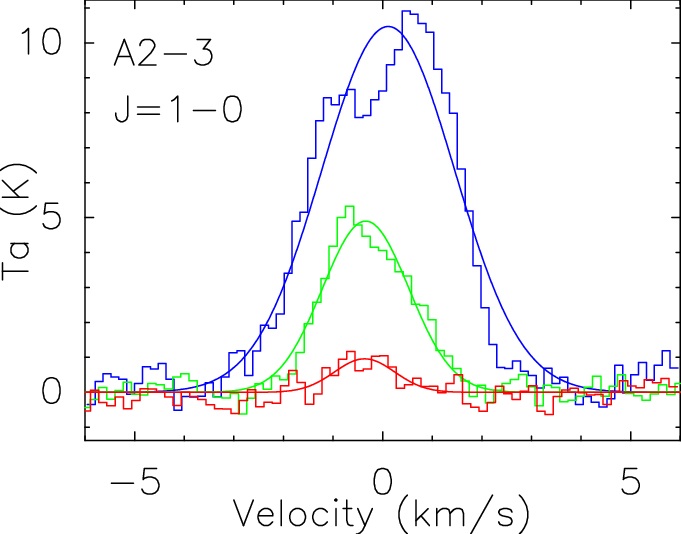}
	\end{minipage}
	\begin{minipage}[c]{0.24\linewidth}
		\centering
		\includegraphics[width=\textwidth]{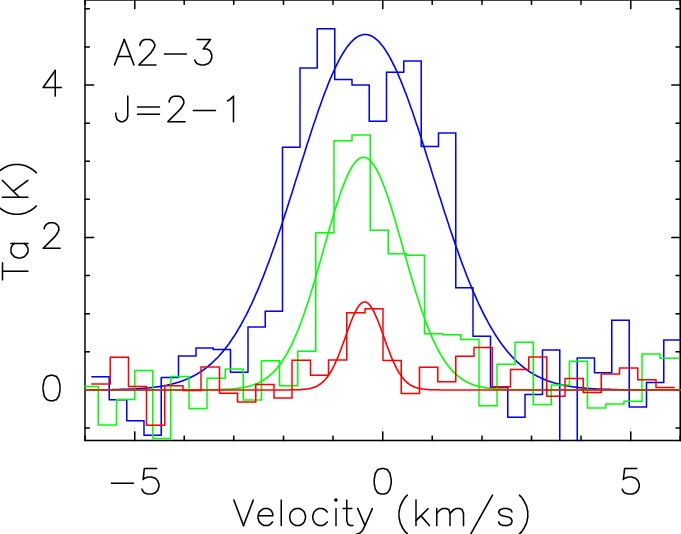}
	\end{minipage}
    	\begin{minipage}[c]{0.24\linewidth}
		\centering
		\includegraphics[width=\textwidth]{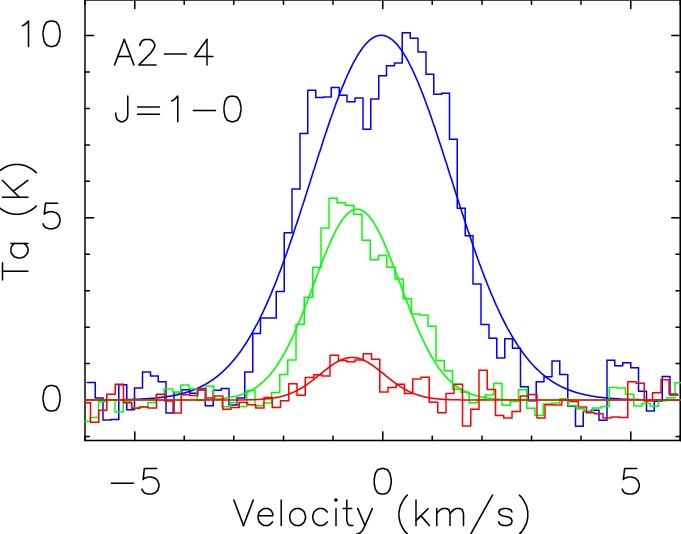}
	\end{minipage}
	\begin{minipage}[c]{0.24\linewidth}
		\centering
		\includegraphics[width=\textwidth]{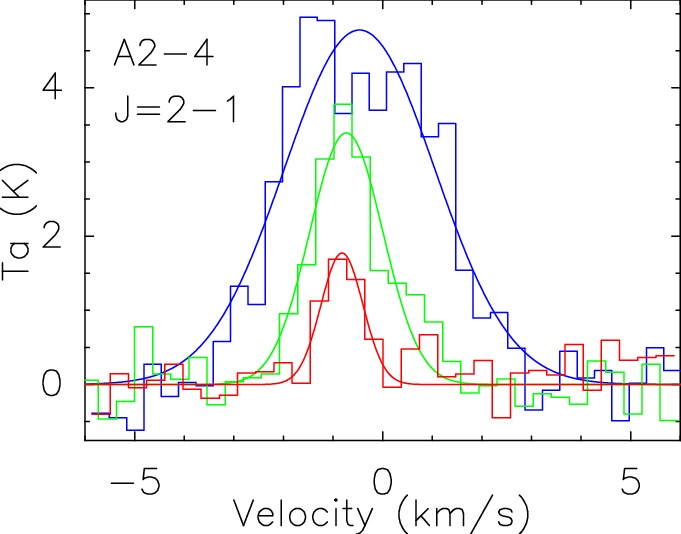}
	\end{minipage}
    	\begin{minipage}[c]{0.24\linewidth}
		\centering
		\includegraphics[width=\textwidth]{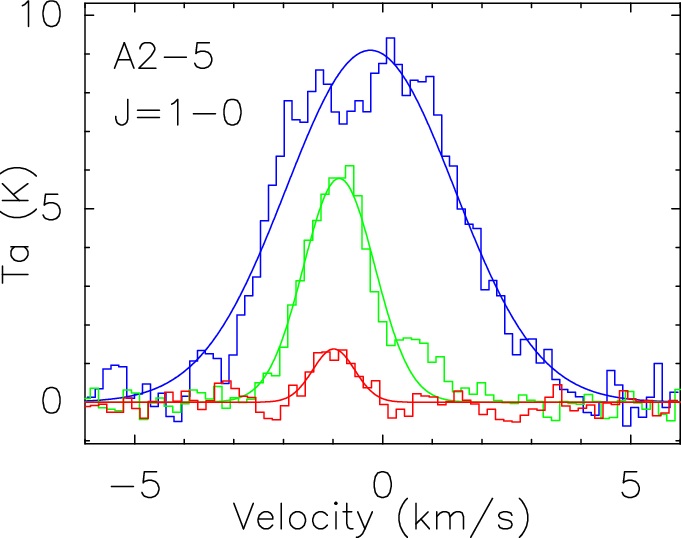}
	\end{minipage}
	\begin{minipage}[c]{0.24\linewidth}
		\centering
		\includegraphics[width=\textwidth]{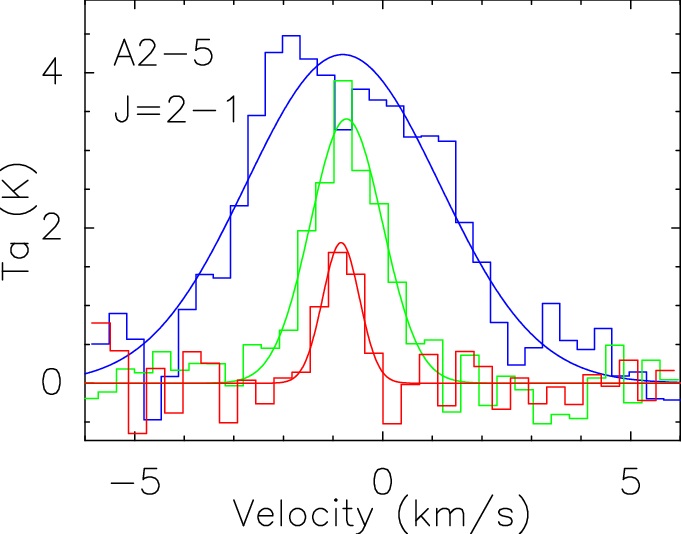}
	\end{minipage}
    	\begin{minipage}[c]{0.24\linewidth}
		\centering
		\includegraphics[width=\textwidth]{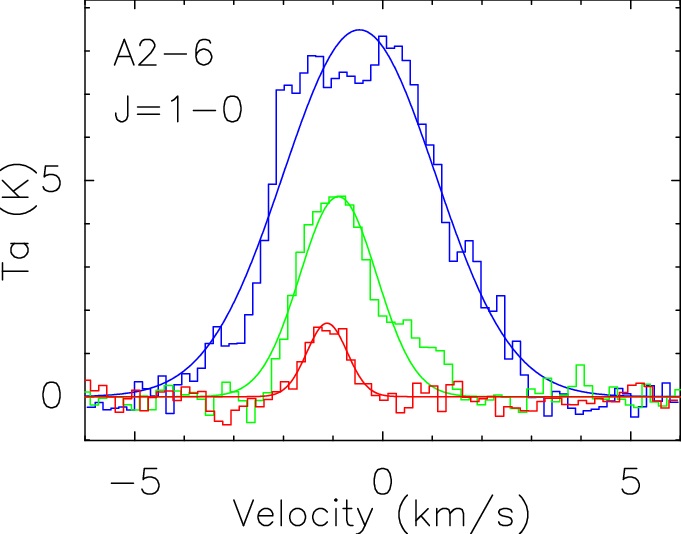}
	\end{minipage}
	\begin{minipage}[c]{0.24\linewidth}
		\centering
		\includegraphics[width=\textwidth]{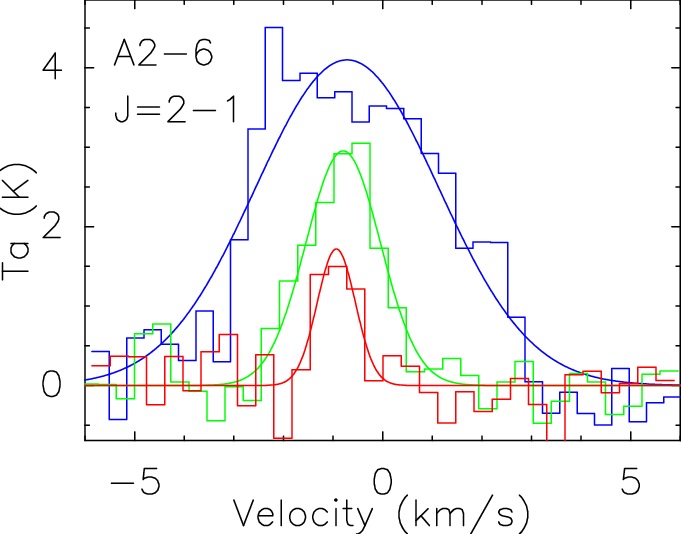}
	\end{minipage}
    	\begin{minipage}[c]{0.24\linewidth}
		\centering
		\includegraphics[width=\textwidth]{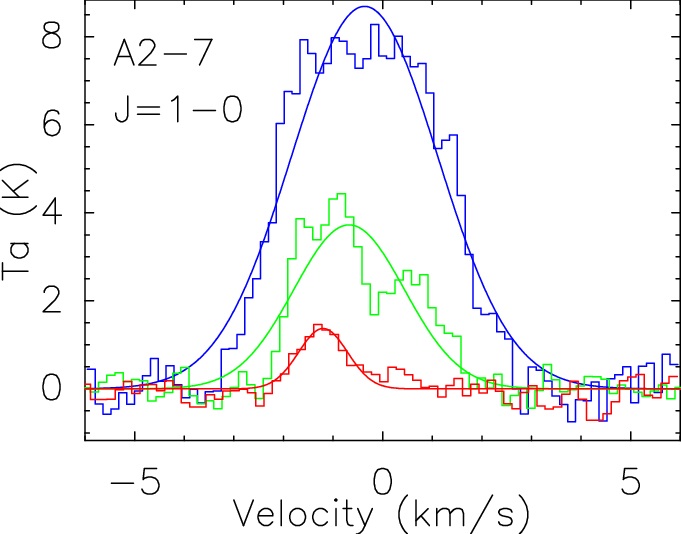}
	\end{minipage}
	\begin{minipage}[c]{0.24\linewidth}
		\centering
		\includegraphics[width=\textwidth]{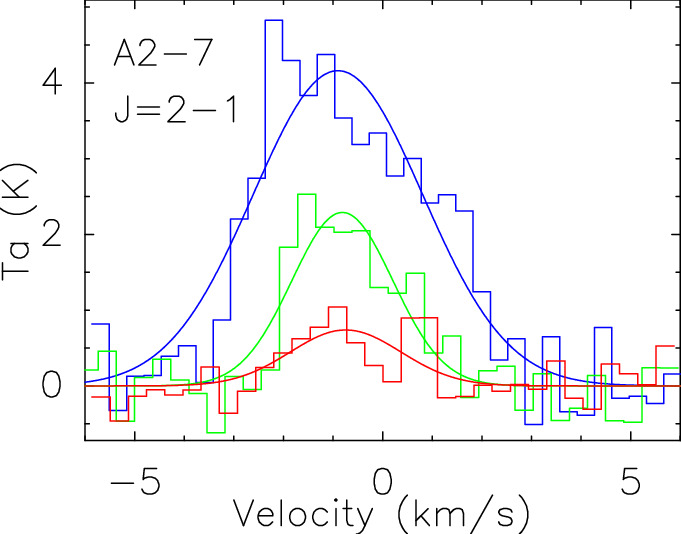}
	\end{minipage}
    	\begin{minipage}[c]{0.24\linewidth}
		\centering
		\includegraphics[width=\textwidth]{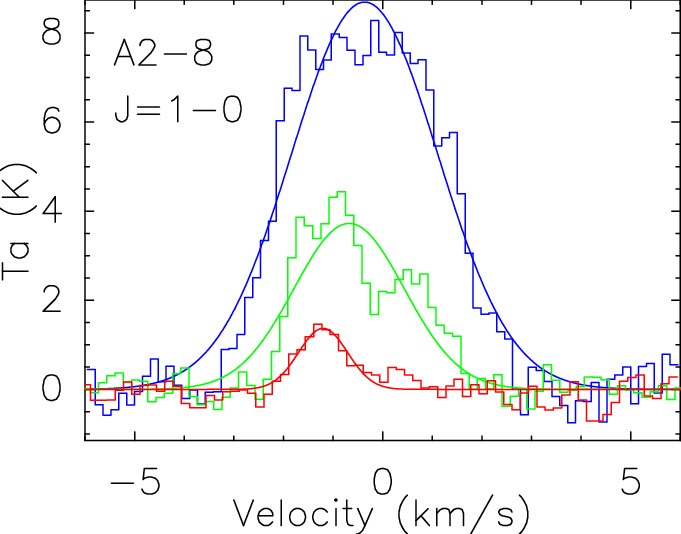}
	\end{minipage}
	\begin{minipage}[c]{0.24\linewidth}
		\centering
		\includegraphics[width=\textwidth]{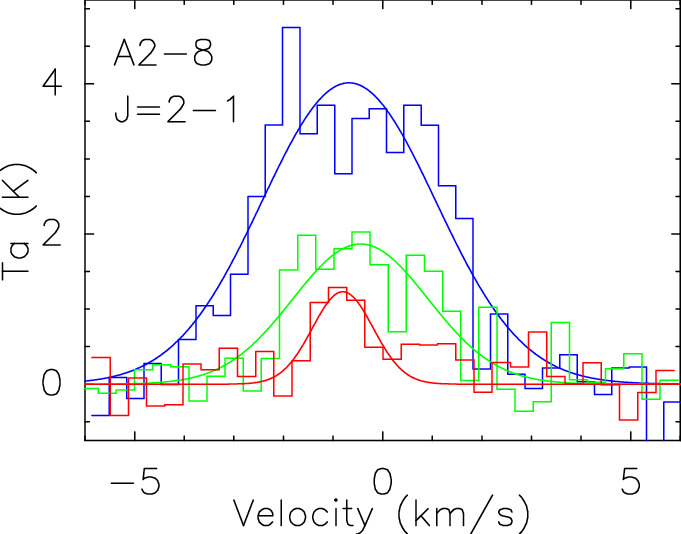}
	\end{minipage}
    	\begin{minipage}[c]{0.24\linewidth}
		\centering
		\includegraphics[width=\textwidth]{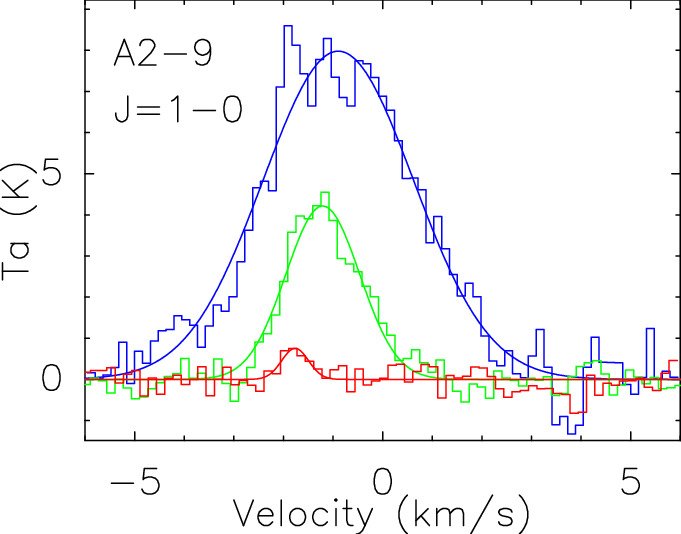}
	\end{minipage}
	\begin{minipage}[c]{0.25\linewidth}
		\centering
		\includegraphics[width=\textwidth]{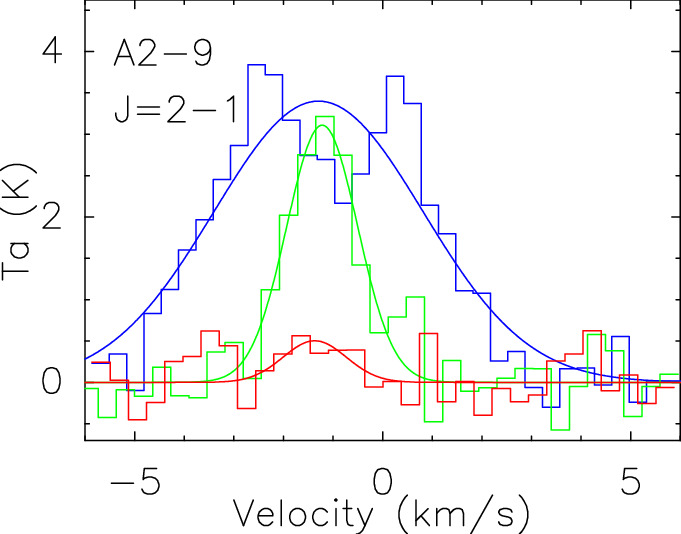}
	\end{minipage}
    	\begin{minipage}[c]{0.25\linewidth}
		\centering
		\includegraphics[width=\textwidth]{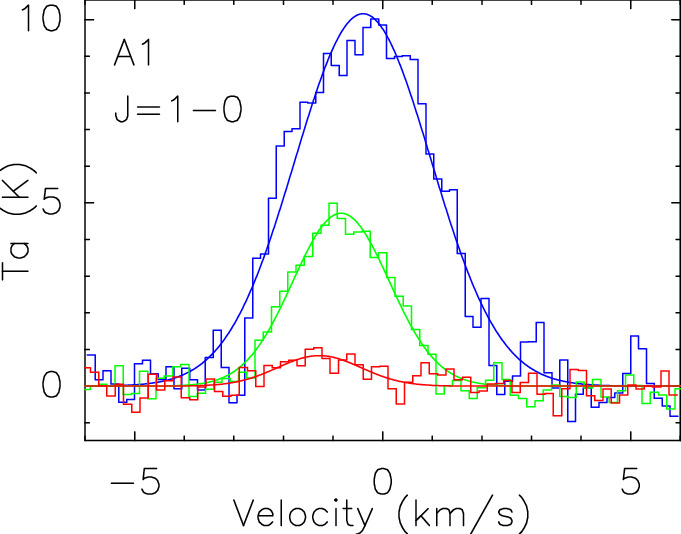}
	\end{minipage}
\caption{CO J=1-0 and J=2-1 spectra at ten continuum peaks, except A1, which was not observed by CO J=2-1. For each plot, the three lines of $^{12}$CO, $^{13}$CO, C$^{18}$O with corresponding single-component gaussian fitting results are colored blue, green, and red, respectively. The velocity ranges from -6~km~s$^{-1}$ to 6~km~s$^{-1}$ for all plots.}
\label{fig:co-spec}
\end{figure*}

\begin{figure*}
	\begin{minipage}[c]{\linewidth}
		\centering
		\includegraphics[width=\textwidth]{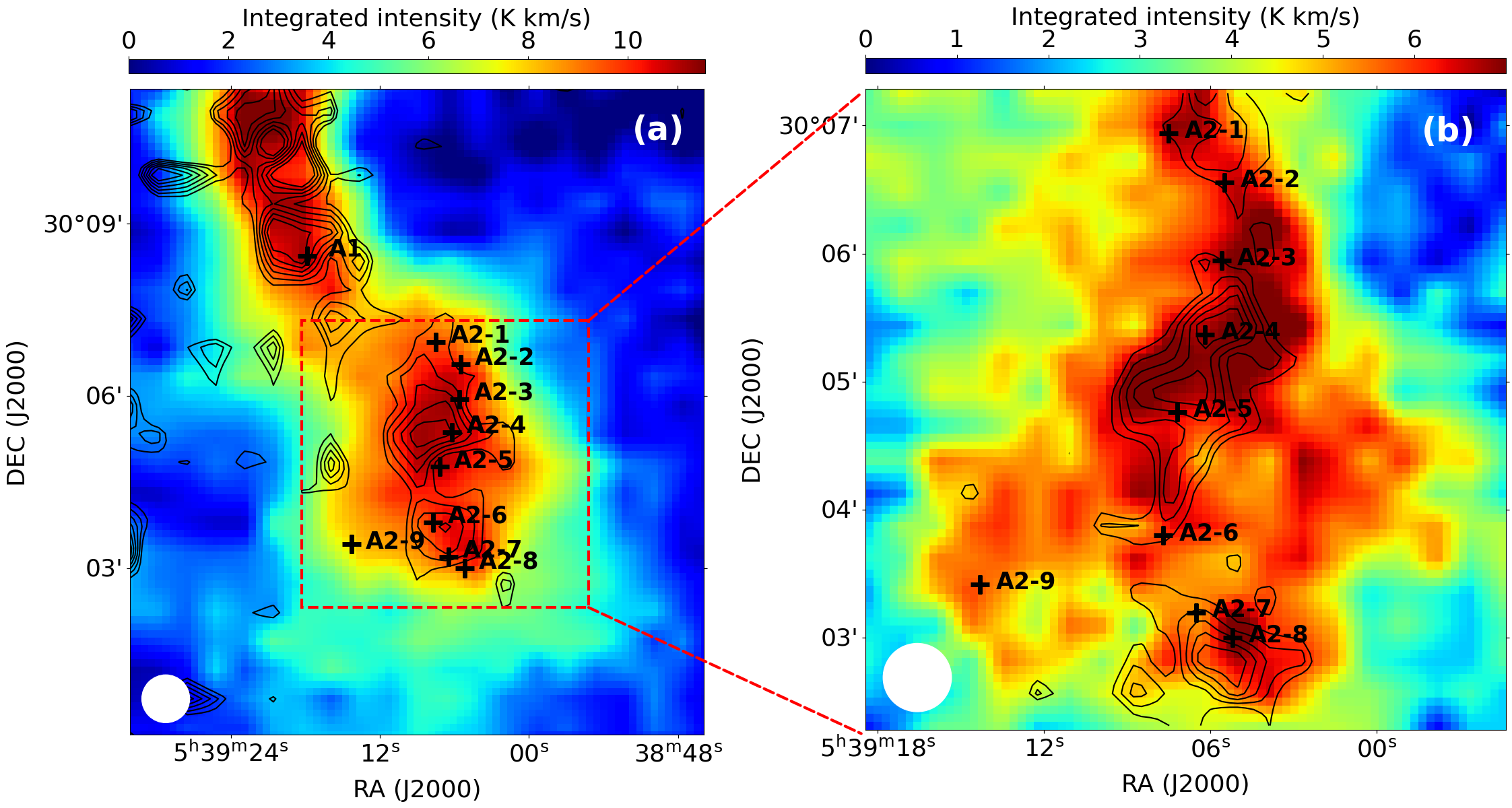}
	\end{minipage}
        \begin{minipage}[c]{\linewidth}
		\centering
		\includegraphics[width=\textwidth]{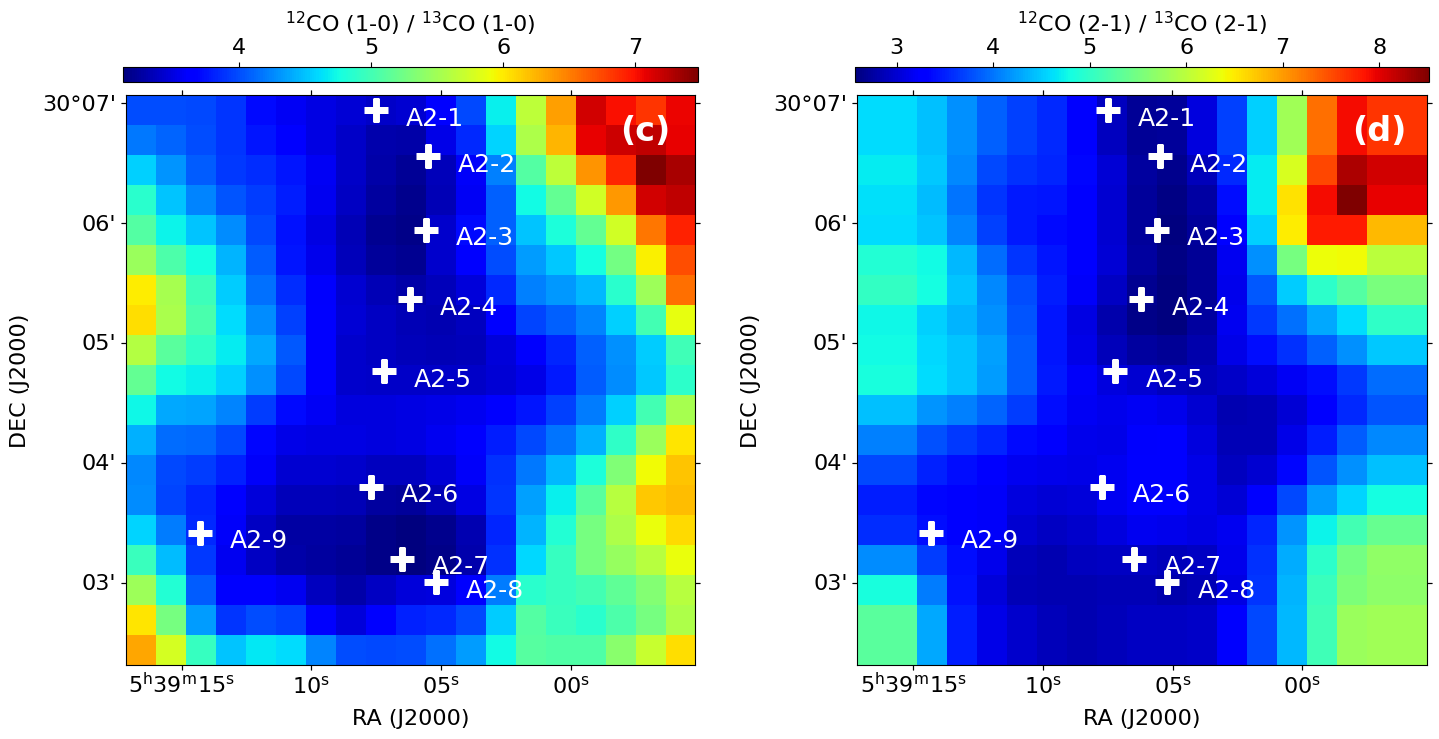}
	\end{minipage}
\caption{(a): The $^{13}$CO (1-0) integrated intensity map is shown in color scale, and C$^{18}$O (1-0) map is shown in black contours at levels of 0.3--1.5 K~km~s$^{-1}$, increasing in steps of 0.2 K~km~s$^{-1}$. The positions of the ten 850 $\mu$m continuum peaks are indicated with crosses and labeled. (b): The $^{13}$CO (2-1) integrated intensity map is shown in color scale, and C$^{18}$O (2-1) map is shown in black contours from 1.4 to 2.2 K~km~s$^{-1}$, in steps of 0.2 K~km~s$^{-1}$. The corresponding beam sizes are plotted as white circle. (c)(d): Integrated intensity ratios of $^{12}$CO (1–0) to $^{13}$CO (1–0) and $^{12}$CO (2–1) to $^{13}$CO (2–1). The 850 $\mu$m continuum peaks are labeled.}
\label{fig:CO}
\end{figure*}

\cite{zha16} previously examined the CO J=1-0 emission in G178 and found two clumps, A1 and A2.  Derived from the average CO spectra of clump A2, the temperature T$_{ex}$ is recorded as 13.0$\pm$1.0 K, with an H$_2$ column density of $(1.4\pm0.2)\times 10^{22}$~cm$^{-2}$, similar to the SED results. In this work, we have extracted the three CO spectra at ten distinct 850 $\mu$m emission peaks. Utilizing the same methodology outlined in \cite{zha16}, the spectra are fitted with a gaussian profile (see Figure \ref{fig:co-spec}), giving the central velocity (V$_{lsr}$), line width (FWHM), and peak brightness temperature (T$_B$) in Table \ref{table:obs-fit}. For each continuum peak position, the excitation temperature (T$_{ex}$), optical depth, and column density for these spectra are calculated and detailed in Table \ref{table:CO}. The $^{12}$CO column densities and optical depths are calculated by adopting the typical ratio [$^{12}$CO]/[$^{13}$CO]=60 in the ISM \citep{deh08}. The calculated excitation temperatures span from 11.4 to 14.3 K for the J=1-0 transitions, again aligning closely with the dust temperatures of the cold component from the SED fit at $\sim$12.4~K, suggesting a good coupling between local dust and gas components. 

For the J=2-1 lines (Figure \ref{fig:moments} (b)), the contours of C$^{18}$O (2-1) exhibit a filamentary and fragmented configuration, demonstrating substantial correspondence with the 850 $\mu$m continuum peaks. In addition, parameters for J=2-1 lines were derived (refer to Table \ref{table:CO}). C$^{18}$O exhibit predominantly optically thin characteristics for both J=2-1 and J=1-0 emissions according to the compact distribution and calculated $\tau_{18}$ in Table \ref{table:CO}. Generally, the resulting $\tau$ shows that J=2-1 are optically thicker than the J=1-0 lines, due to its greater absorption coefficient and higher resolution that traces the higher density part of the cloud. We notice the T$_{ex}$ of J=2-1 lines are 3--4~K lower than the J=1-0 lines, which could be explained by the inner collapsing cores colder than the gas of filament. Another explanation is that the T$_{ex}$ of J=2-1 lines are underestimated. For example, towards the lines of A2-9 in Figure A1, the $^{12}$CO (2-1) spectrum shows a dip corresponding to the peak of $^{13}$CO, indicating feature of self-absorption. Therefore, the peak intensity of $^{12}$CO (2-1) is underestimated by the single-component gaussian fit, which leads to a lower T$_{ex}$. Hence, the T$_{ex}$ of J=1-0 lines are used in the following discussion.

To analyze the optical depth throughout the region, the integrated intensity ratios of $^{12}$CO and $^{13}$CO were calculated for the J=2-1 and J=1-0 lines, shown in color images in panels (c) and (d) of Figure \ref{fig:CO}. Smaller ratios indicate higher optical depths for CO emission \citep{liu16}, indicating that the CO emission at the clump edge is less optically thick than that at the clump center.

The fractional abundances of the molecules were determined by dividing their molecular column densities by the H$_2$ column density obtained from the dust SED, $R_{mol}=N_{mol}/N_{H_2}$, as presented in Table \ref{table:CO}. Previous research has indicated that $R_{^{13}CO}$ falls within the range of 1.0--1.7$\times$10$^{-6}$ \citep{mcc80,fre82,lan90,she07,she08,men13}, which is generally in agreement with our findings (0.9--2.3$\times$10$^{-6}$ for J=1-0). Furthermore, \cite{fre82} asserted that $R_{C^{18}O}$ is approximately 1.4$\times$10$^{-7}$ in the $\rho$ Oph and Taurus molecular cloud complexes, which is also roughly consistent with our results (0.7--2.5$\times$10$^{-7}$ for J=1-0).

The abundances exhibit variation along the filament. Locations exhibiting a stronger continuum are generally associated with increased fractional abundances of both $^{13}$CO and C$^{18}$O. Nonetheless, the capacity of the CO fractional abundance to indicate evolutionary stages remains an open question. The column density ratios between $^{13}$CO and C$^{18}$O, as presented in the last column of Table \ref{table:CO}, demonstrate a median value of 9.1. This value slightly exceeds those observed in YSOs/HII regions \citep[7–8;][]{are18} and in regions of high mass star formation \citep[5;][]{par18}.

\begin{table*}
   \caption{Observed parameters of CO and dense molecular lines}\label{table:obs-fit}
   \centering
   \begin{tabular}{c|ccc|ccc|ccc}
\hline \hline
      &  \multicolumn{3}{c|}{$^{12}$CO J=1-0} & \multicolumn{3}{c|}{$^{13}$CO J=1-0} & \multicolumn{3}{c}{C$^{18}$O J=1-0} \\
 Name &  V$_{lsr}$  & FWHM  & T$_{B}$ &  V$_{lsr}$  & FWHM  & T$_{B}$ &  V$_{lsr}$  & FWHM  & T$_B$ \\
    &(km~s$^{-1}$) &(km~s$^{-1}$) &(K)&(km~s$^{-1}$) &(km~s$^{-1}$) &(K)&(km~s$^{-1}$) &(km~s$^{-1}$) &(K) \\
\hline
%\multirow{10}{*}{\rotatebox{90}{J=1-0}} 
A1&-0.39(0.04)&3.23(0.08)&10.16(0.94)&-0.84(0.03)&2.29(0.07)&4.72(0.22)&-1.28(0.17)&1.98(0.41)&0.83(0.18) \\
A2-1&0.01(0.02)&2.93(0.05)&10.88(4.57)&-0.30(0.02)&1.95(0.05)&4.56(0.34)&-0.49(0.07)&0.94(0.17)&1.07(0.22) \\
A2-2&0.19(0.02)&2.88(0.05)&10.42(1.28)&-0.23(0.02)&1.92(0.05)&4.97(0.47)&-0.38(0.11)&1.12(0.29)&0.85(0.30) \\
A2-3&0.11(0.02)&3.16(0.05)&10.47(2.28)&-0.34(0.02)&2.05(0.05)&4.90(0.34)&-0.36(0.10)&1.43(0.24)&0.96(0.29) \\
A2-4&-0.03(0.03)&3.33(0.06)&10.01(9.03)&-0.52(0.02)&2.06(0.05)&5.24(0.24)&-0.62(0.08)&1.51(0.19)&1.17(0.19) \\
A2-5&-0.28(0.03)&3.56(0.06)&9.76(0.92)&-0.70(0.02)&2.04(0.05)&5.15(0.20)&-0.97(0.12)&1.93(0.27)&0.92(0.16) \\
A2-6&-0.47(0.03)&3.61(0.06)&8.49(0.52)&-0.89(0.02)&1.83(0.05)&4.63(0.15)&-1.12(0.04)&0.96(0.09)&1.70(0.15) \\
A2-7&-0.36(0.03)&3.48(0.06)&8.69(0.73)&-0.68(0.03)&2.62(0.07)&3.73(0.19)&-1.20(0.07)&1.16(0.16)&1.37(0.17) \\
A2-8&-0.36(0.03)&3.48(0.06)&8.69(0.73)&-0.68(0.03)&2.62(0.07)&3.73(0.19)&-1.20(0.07)&1.16(0.16)&1.37(0.17) \\
A2-9&-0.89(0.03)&3.57(0.08)&7.98(0.35)&-1.21(0.03)&1.75(0.06)&4.21(0.15)&-1.76(0.10)&0.64(0.23)&0.75(0.21) \\
\hline \hline
      &  \multicolumn{3}{c|}{$^{12}$CO J=2-1} & \multicolumn{3}{c|}{$^{13}$CO J=2-1} & \multicolumn{3}{c}{C$^{18}$O J=2-1} \\
 Name &  V$_{lsr}$  & FWHM  & T$_{B}$ &  V$_{lsr}$  & FWHM  & T$_{B}$ &  V$_{lsr}$  & FWHM  & T$_B$ \\
    &(km~s$^{-1}$) &(km~s$^{-1}$) &(K)&(km~s$^{-1}$) &(km~s$^{-1}$) &(K)&(km~s$^{-1}$) &(km~s$^{-1}$) &(K) \\
\hline
%\multirow{9}{*}{\rotatebox{90}{J=2-1}}
A2-1&-0.27(0.05)&3.16(0.10)&5.14(0.93)&-0.31(0.07)&1.93(0.16)&3.32(0.74)&-0.55(0.07)&0.68(0.14)&1.42(0.34) \\
A2-2&-0.03(0.05)&3.10(0.11)&4.91(0.95)&-0.24(0.06)&2.05(0.13)&2.92(0.71)&-0.55(0.16)&1.21(0.44)&0.98(0.39) \\
A2-3&-0.35(0.06)&3.20(0.13)&4.66(0.82)&-0.38(0.07)&1.89(0.17)&3.05(0.58)&-0.36(0.09)&0.90(0.24)&1.15(0.37) \\
A2-4&-0.47(0.06)&3.58(0.13)&4.78(0.64)&-0.73(0.05)&1.74(0.14)&3.39(0.31)&-0.82(0.07)&1.00(0.17)&1.77(0.30) \\
A2-5&-0.80(0.07)&4.56(0.16)&4.25(0.41)&-0.73(0.06)&1.71(0.14)&3.40(0.35)&-0.84(0.06)&0.87(0.13)&1.81(0.29) \\
A2-6&-0.72(0.07)&4.36(0.15)&4.10(0.44)&-0.79(0.05)&1.77(0.13)&2.95(0.27)&-0.93(0.07)&0.91(0.13)&1.72(0.28) \\
A2-7&-0.90(0.08)&4.03(0.17)&4.16(0.38)&-0.81(0.10)&2.40(0.22)&2.29(0.34)&-0.75(0.27)&2.63(0.52)&0.74(0.30) \\
A2-8&-0.68(0.07)&4.13(0.15)&4.01(0.45)&-0.45(0.14)&3.22(0.30)&1.87(0.60)&-0.80(0.20)&1.44(0.99)&1.23(0.46) \\
A2-9&-1.29(0.09)&4.96(0.19)&3.40(0.28)&-1.22(0.05)&1.68(0.13)&3.11(0.23)&-1.37(0.31)&1.50(0.66)&0.50(0.23) \\
\hline \hline
      &  \multicolumn{3}{c|}{N$_{2}$H$^{+}$} & \multicolumn{3}{c|}{HCO$^+$} & \multicolumn{3}{c}{H$_2$CO} \\
 Name &  V$_{lsr}$  & FWHM  & T$_{B}$ &  V$_{lsr}$  & FWHM  & T$_{B}$ &  V$_{lsr}$  & FWHM  & T$_B$ \\
      &(km~s$^{-1}$) &(km~s$^{-1}$) &(K)&(km~s$^{-1}$) &(km~s$^{-1}$) &(K)&(km~s$^{-1}$) &(km~s$^{-1}$) &(K) \\
\hline
A1&-2.99(0.04)&0.43(0.11)&0.22(0.14)&-1.94(0.03)&1.00(0.08)&1.55(0.09)&-1.54(0.12)&2.07(0.30)&0.44(0.03) \\
A2-1&-0.53(0.03)&0.56(0.08)&0.44(0.16)&-0.65(0.03)&1.22(0.08)&1.55(0.08)&-0.57(0.04)&1.30(0.11)&0.71(0.04) \\
A2-2&-0.49(0.03)&0.53(0.06)&0.39(0.10)&-0.72(0.03)&1.14(0.06)&1.71(0.08)&-0.47(0.03)&1.23(0.07)&0.93(0.04) \\
A2-3&-0.61(0.02)&0.55(0.04)&0.75(0.13)&-0.94(0.02)&1.01(0.06)&2.03(0.09)&-0.61(0.04)&1.45(0.10)&0.77(0.03) \\
A2-4&-0.84(0.02)&0.63(0.04)&0.91(0.13)&-1.16(0.02)&1.10(0.06)&2.19(0.09)&-0.90(0.03)&1.36(0.07)&1.03(0.04) \\
A2-5&-0.98(0.02)&0.56(0.05)&0.69(0.14)&-1.00(0.07)&2.16(0.20)&1.04(0.07)&-0.83(0.03)&1.29(0.08)&0.98(0.07) \\
A2-6&-0.95(0.03)&0.76(0.06)&0.68(0.05)&-1.15(0.07)&1.82(0.15)&0.96(0.07)&-0.90(0.04)&1.38(0.09)&0.75(0.06) \\
A2-7&--&--&--&-1.35(0.07)&1.50(0.17)&0.76(0.07)&-1.21(0.06)&1.29(0.13)&0.65(0.06) \\
A2-8&-1.45(0.05)&0.62(0.11)&0.34(0.05)&-1.15(0.09)&1.73(0.21)&0.61(0.06)&-1.31(0.07)&1.16(0.15)&0.58(0.06) \\
A2-9&-1.24(0.03)&0.66(0.07)&0.51(0.04)&-1.15(0.09)&2.06(0.22)&0.68(0.06)&-1.14(0.04)&1.18(0.11)&0.81(0.07) \\
\hline
   \end{tabular}
\tablenotetext{}{Note: The central velocity (V$_{lsr}$), line width (FWHM), and peak brightness temperature (T$_B$) are derived with a gaussian fitting for each spectrum.}
\end{table*}

\begin{deluxetable*}{c|ccccccccccc}
	\tabletypesize{\scriptsize}
	\tablewidth{0pt}
	\setlength{\tabcolsep}{0.05in}
	\tablecaption{Parameters of CO isotopologue emission \label{table:CO}}
	\tablehead{
J & Name& T$_{ex}$ & $\tau_{12}$ & N$_{^{12}CO}$ & $\tau_{13}$ & N$_{^{13}CO}$ & $R_{^{13}CO}$ & $\tau_{18}$ & N$_{C^{18}O}$ & $R_{C^{18}O}$ & N$_{^{13}CO}$/N$_{C{^{18}O}}$ \\
& & K &  & $10^{17}$~cm$^{-2}$ & & $10^{16}$~cm$^{-2}$ & $10^{-6}$ & & $10^{15}$~cm$^{-2}$ & $10^{-7}$ & } 
\startdata
&A1&   13.6(1.0)& 37(5)& 9.3(1.9)& 0.62(0.09)& 1.6(0.3)&	2.3(0.5)	& 0.09(0.02)& 1.7(0.6)&	2.5(0.9)	&	9(4)	\\
&A2-1& 14.3(1.0)& 32(19)& 7.6(4.4)& 0.54(0.31)& 1.3(0.7)&	1.4(0.8)	& 0.10(0.05)& 1.1(0.6)&	1.2(0.7)	&	12(10)	\\
&A2-2& 13.8(1.0)& 39(8)& 8.4(2.2)& 0.65(0.14)& 1.4(0.4)&	0.9(0.2)	& 0.09(0.03)& 1.0(0.5)&	0.7(0.3)	&	14(8)	\\
&A2-3& 13.9(1.0)& 38(12)& 8.8(3.0)& 0.63(0.20)& 1.5(0.5)&	1.1(0.4)	& 0.10(0.04)& 1.4(0.6)&	1.1(0.5)	&	10(6)	\\
J=1-0&A2-4& 13.4(1.0)& 44(5)& 9.7(6.4)& 0.74(0.09)& 1.6(1.1)&1.3(0.9)	& 0.12(0.12)& 1.9(0.4)&	1.5(0.3)	&	9(6)	\\
&A2-5& 13.2(1.0)& 45(7)& 9.4(1.9)& 0.75(0.11)& 1.6(0.3)&	1.8(0.4)	& 0.10(0.02)& 1.8(0.5)&	2.0(0.6)	&	9(3)	\\
&A2-6& 11.9(1.0)& 47(5)& 7.4(1.4)& 0.79(0.08)& 1.2(0.2)&	1.2(0.2)	& 0.22(0.03)& 1.7(0.4)&	1.6(0.3)	&	7(2)	\\
&A2-7& 12.1(1.0)& 34(4)& 7.8(1.6)& 0.56(0.07)& 1.3(0.3)&	0.9(0.2)	& 0.17(0.03)& 1.6(0.4)&	1.2(0.3)	&	8(3)	\\
&A2-8& 12.1(1.0)& 34(4)& 7.8(1.6)& 0.56(0.07)& 1.3(0.3)&	1.6(0.3)	& 0.17(0.03)& 1.6(0.4)&	2.0(0.5)	&	8(3)	\\
&A2-9& 11.4(1.0)& 45(4)& 6.3(1.1)& 0.75(0.06)& 1.1(0.2)&	1.7(0.3)	& 0.10(0.03)& 0.5(0.2)&	0.8(0.4)	&	22(12)	\\
\hline								
&A2-1&  9.9(1.1)&  62(31)&  6.7(2.2)& 1.0(0.5)&  1.1(0.4)&	1.3(0.4)	&  0.32(0.11)&  0.7(0.2)&	0.8(0.2)	&	15(7)	\\
&A2-2&  9.6(1.1)&  54(28)&  5.7(2.0)& 0.9(0.5)&  1.0(0.3)&	0.6(0.2)	&  0.22(0.11)&  0.9(0.4)&	0.6(0.3)	&	11(6)	\\
&A2-3&  9.3(1.1)&  64(29)&  6.4(1.9)& 1.1(0.5)&  1.1(0.3)&	0.8(0.2)	&  0.28(0.12)&  0.8(0.3)&	0.6(0.2)	&	13(6)	\\
&A2-4&  9.5(1.1)&  74(24)&  7.5(1.5)& 1.2(0.4)&  1.3(0.3)&	1.0(0.2)	&  0.46(0.13)&  1.5(0.3)&	1.2(0.3)	&	8(2)	\\
J=2-1&A2-5&  8.9(1.1)&97(34)&  10.4(2.0)& 1.6(0.6)&1.7(0.3)&2.0(0.4)	&  0.55(0.14)&  1.5(0.3)&	1.7(0.3)	&	12(3)	\\
&A2-6&  8.7(1.1)&  76(22)&  7.4(1.3)& 1.3(0.4)&  1.2(0.2)&	1.2(0.2)	&  0.54(0.14)&  1.5(0.3)&	1.4(0.3)	&	8(2)	\\
&A2-7&  8.8(1.1)&  48(13)&  5.2(1.1)& 0.8(0.2)&  0.9(0.2)&	0.6(0.1)	&  0.20(0.09)&  1.6(0.7)&	1.1(0.5)	&	6(3)	\\
&A2-8&  8.6(1.2)&  38(18)&  5.0(1.9)& 0.6(0.3)&  0.8(0.3)&	1.0(0.4)	&  0.37(0.17)&  1.6(1.1)&	1.9(1.3)	&	5(4)	\\
&A2-9&  7.9(1.2)&  148(71)&  19.2(3.5)& 2.5(1.2)&  3.2(0.6)&5.1(1.0)	&  0.16(0.08)&  0.7(0.4)&	1.1(0.6)	&	48(30)	\\
\enddata
\tablenotetext{}{Note: The optical depths of $^{12}$CO, $^{13}$CO, and C$^{18}$O. $\tau_{12}$ = 60 $\tau_{13}$, following \cite{deh08}. R$_{mol}$ is the ratio of the column density between a certain molecule and H$_2$ listed in Table \ref{table:SED}.}
\end{deluxetable*}

\begin{deluxetable*}{ccccccccccc}
	\setlength{\tabcolsep}{0.05in}
	\tabletypesize{\scriptsize}
	\tablewidth{0pt}
	\tablecaption{Optical depth, column density and abundance ratio of three dense gas tracers \label{table:lines}}
	\tablehead{
Name& $\tau_{N_{2}H^{+}}$ & N$_{N_{2}H^{+}}$ & $R_{N_{2}H^{+}}$ & $\tau_{HCO^+}$ & N$_{HCO^+}$ &$R_{HCO^+}$& $\tau_{H_2CO}$ & N$_{H_2CO}$ & $R_{H_2CO}$ & N$_{N_{2}H^{+}}$/N$_{HCO^+}$\\ 
		& & $10^{13}$~cm$^{-2}$ & $10^{-9}$ & & $10^{12}$~cm$^{-2}$ & $10^{-10}$ & & $10^{13}$~cm$^{-2}$ & $10^{-9}$ & }
	\startdata
A1		&0.49(0.42)&0.85(0.22)&0.66(0.17)&0.16(0.02)&2.25(0.13)&1.74(0.11)&0.05(0.01)&1.64(0.17)&1.27(0.14)&3.8(1.0)\\
A2-1	&0.17(0.16)&2.12(0.36)&1.22(0.21)&0.15(0.01)&2.80(0.15)&1.61(0.09)&0.07(0.01)&1.68(0.11)&0.97(0.07)&7.6(1.3)\\
A2-2	&0.58(0.18)&1.83(0.27)&1.07(0.16)&0.17(0.01)&2.86(0.13)&1.67(0.08)&0.10(0.01)&2.08(0.10)&1.22(0.06)&6.4(1.0)\\
A2-3	&0.16(0.08)&3.61(0.45)&1.82(0.23)&0.21(0.02)&3.08(0.14)&1.56(0.07)&0.08(0.01)&2.03(0.12)&1.03(0.06)&11.7(1.5)\\
A2-4	&0.17(0.06)&5.09(0.60)&1.88(0.22)&0.24(0.02)&3.60(0.15)&1.33(0.06)&0.11(0.01)&2.56(0.12)&0.94(0.04)&14.1(1.8)\\
A2-5	&0.11(0.09)&3.46(0.47)&1.51(0.21)&0.11(0.01)&3.11(0.21)&1.35(0.09)&0.11(0.01)&2.32(0.12)&1.01(0.05)&11.1(1.7)\\
A2-6	&0.10(0.09)&4.89(0.63)&2.16(0.28)&0.12(0.01)&2.32(0.17)&1.03(0.08)&0.10(0.01)&1.96(0.12)&0.87(0.05)&21.1(3.1)\\
A2-7	&--&--&--&0.09(0.01)&1.50(0.14)&0.68(0.06)&0.08(0.01)&1.59(0.14)&0.73(0.06)&--\\
A2-8	&0.10(0.05)&1.98(0.41)&0.99(0.20)&0.07(0.01)&1.39(0.14)&0.70(0.07)&0.07(0.01)&1.27(0.13)&0.64(0.07)&14.2(3.3)\\
A2-9	&0.10(0.06)&3.25(0.47)&2.60(0.38)&0.09(0.01)&1.80(0.16)&1.44(0.13)&0.11(0.01)&1.85(0.13)&1.48(0.11)&18.1(3.0)\\
	\enddata
\tablenotetext{}{Note: R$_{mol}$ is the ratio of the column density between a certain molecule and H$_2$ listed in Table \ref{table:SED}.}
\end{deluxetable*}

\begin{figure*}
	\begin{minipage}[c]{0.24\linewidth}
		\centering
		\includegraphics[width=\textwidth]{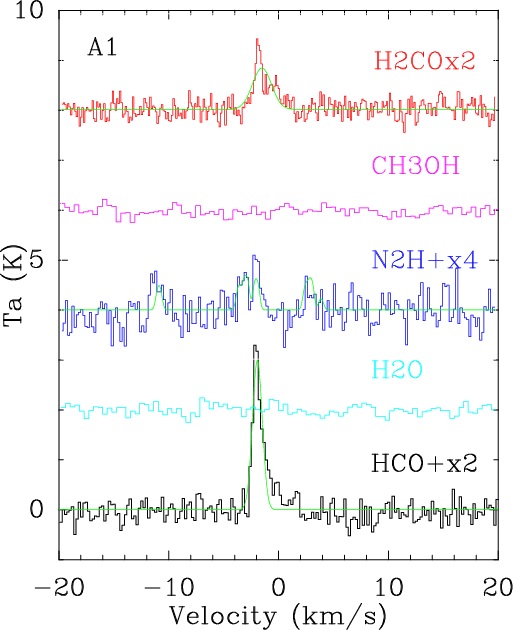}
	\end{minipage}
	\begin{minipage}[c]{0.24\linewidth}
		\centering
		\includegraphics[width=\textwidth]{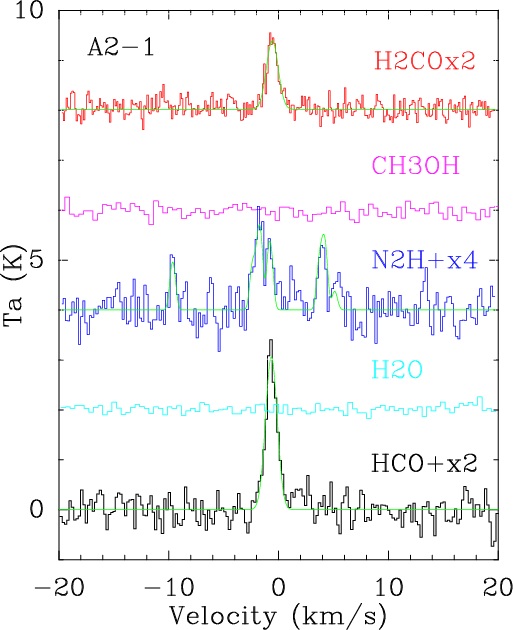}
	\end{minipage}
    	\begin{minipage}[c]{0.24\linewidth}
		\centering
		\includegraphics[width=\textwidth]{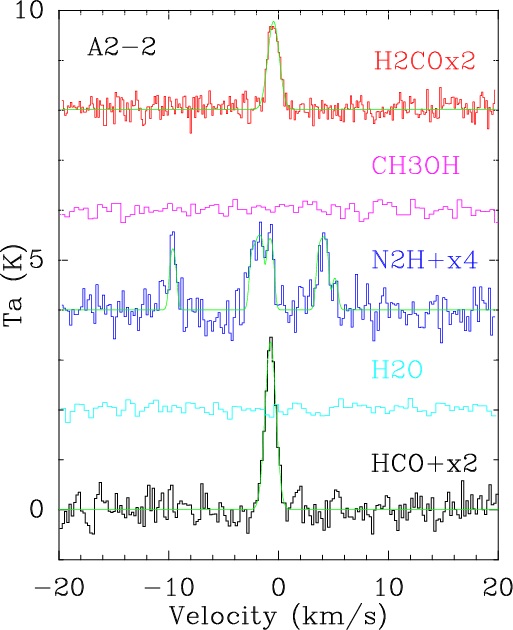}
	\end{minipage}
	\begin{minipage}[c]{0.24\linewidth}
		\centering
		\includegraphics[width=\textwidth]{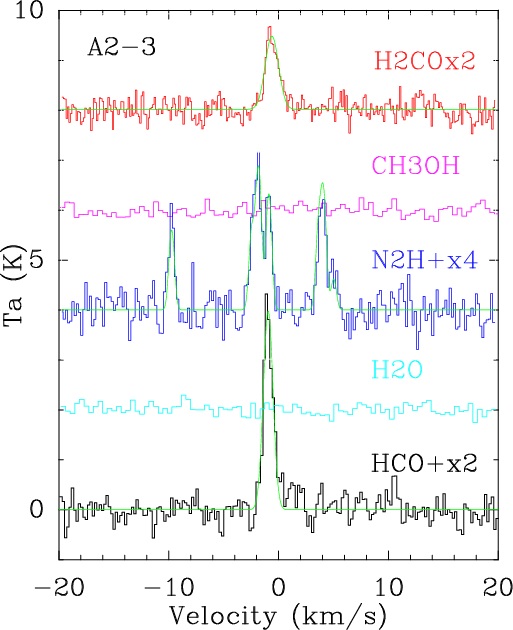}
	\end{minipage}
    	\begin{minipage}[c]{0.24\linewidth}
		\centering
		\includegraphics[width=\textwidth]{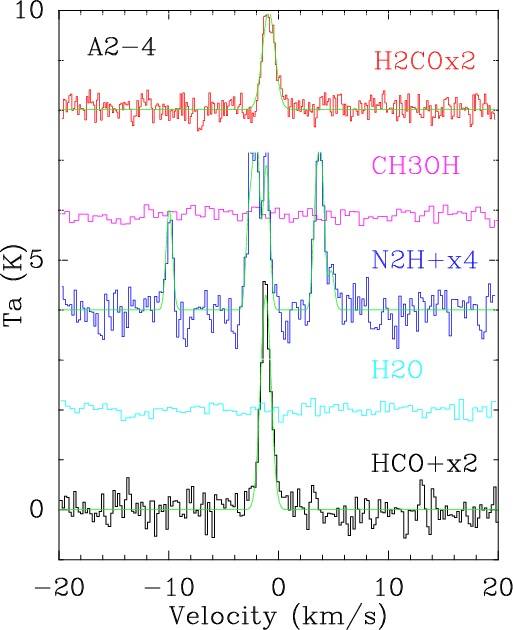}
	\end{minipage}
	\begin{minipage}[c]{0.24\linewidth}
		\centering
		\includegraphics[width=\textwidth]{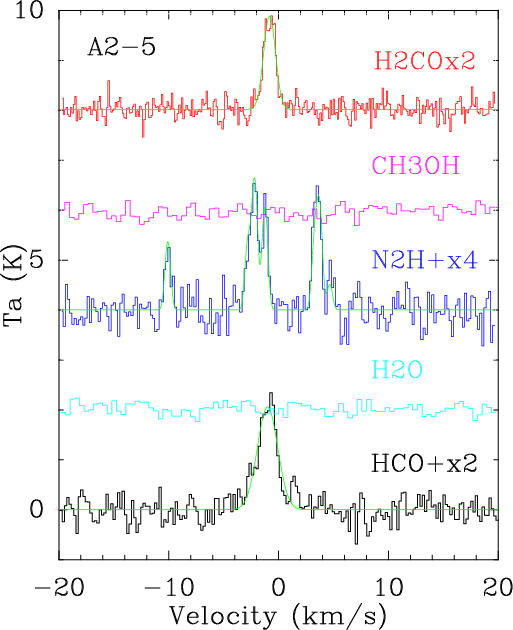}
	\end{minipage}
    	\begin{minipage}[c]{0.24\linewidth}
		\centering
		\includegraphics[width=\textwidth]{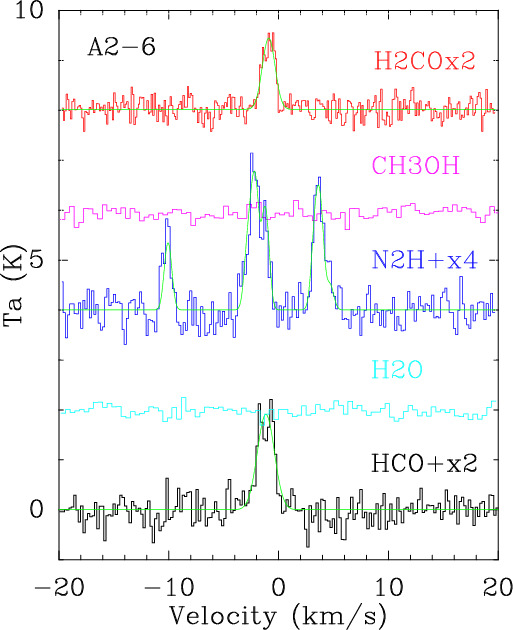}
	\end{minipage}
	\begin{minipage}[c]{0.24\linewidth}
		\centering
		\includegraphics[width=\textwidth]{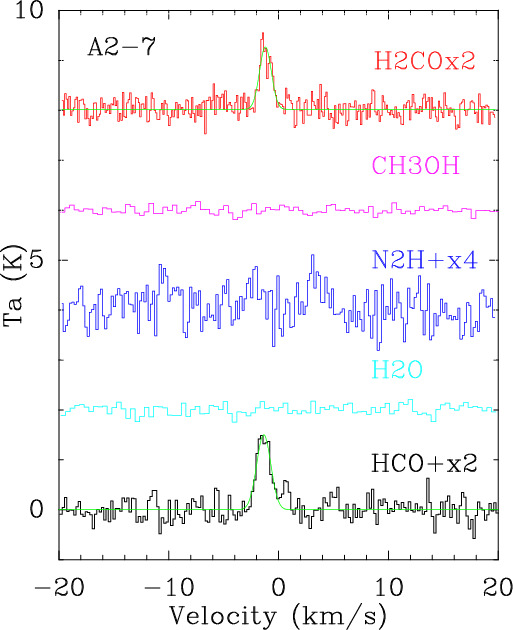}
	\end{minipage}
    	\begin{minipage}[c]{0.24\linewidth}
		\centering
		\includegraphics[width=\textwidth]{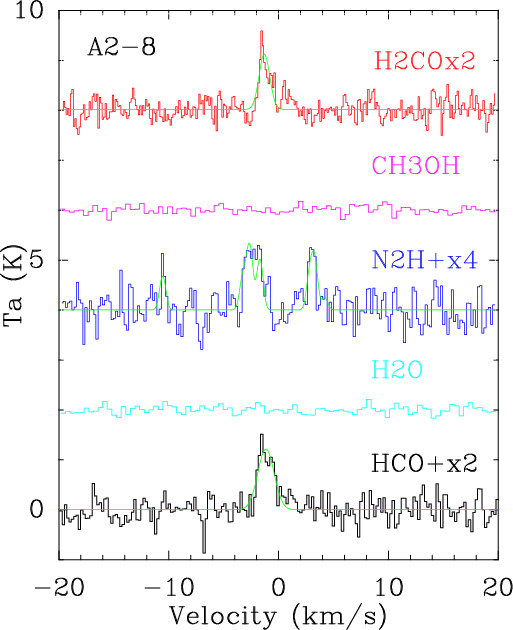}
	\end{minipage}
	\begin{minipage}[c]{0.24\linewidth}
		\centering
		\includegraphics[width=\textwidth]{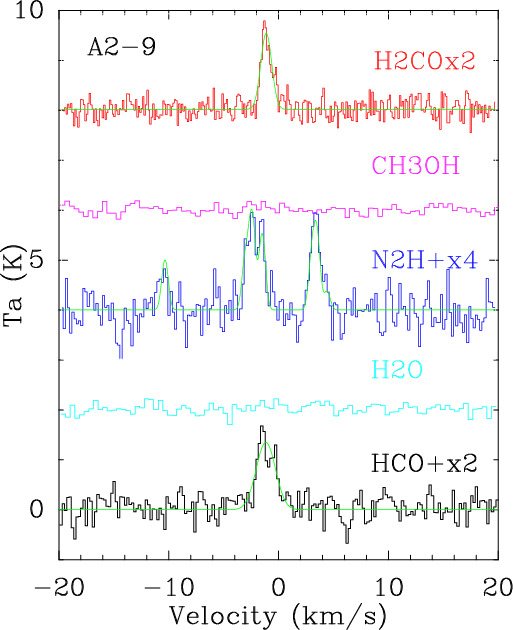}
	\end{minipage}
\caption{Molecular spectra at ten continuum peaks. From bottom to top of each subplot, the spectra of HCO$^+$, H$_2$O, N$_2$H$^+$, CH$_3$OH, and H$_2$CO are added by 0, 2, 4, 6, 8 K, respectively. The intensity of H$_2$CO, N$_2$H$^+$, and HCO$^+$ are multiplied by a factor of two, four, and two, respectively. The green lines show one component fitting for H$_2$CO, HCO$^+$, and hyperfine structure fitting for N$_2$H$^+$. No detections of CH$_3$OH and H$_2$O are found for ten peaks.}
\label{fig:spectrum}
\end{figure*}

\subsubsection{N$_2$H$^+$ J=1-0 hyperfine structure}
The N$_2$H$^+$ emission observed with the KVN traces the dense gas of G178. 
Among the ten continuum peaks analyzed, nine (with the exception of A2-7) display N$_2$H$^+$ emission, with the main component value exceeding 1--5 times the noise level ($\sim$0.2~K). The detected N$_2$H$^+$ spectra are presented in blue in Figure \ref{fig:spectrum}. The GILDAS package's hyperfine structure fitting method \footnote{\href{http://www.iram.fr/IRAMFR/GILDAS}{http://www.iram.fr/IRAMFR/GILDAS}} was employed to directly obtain the parameters of V$_{lsr}$, FWHM, and peak intensity (T$_B$). The fitting results are shown in green lines in Figure \ref{fig:spectrum}, and listed in Table \ref{table:obs-fit}.
Assuming local thermodynamic equilibrium (LTE) conditions, the total column density, N, can be derived from \citep{gar91}:
\begin{eqnarray}
N=\frac{3h}{8\pi^3\mu^2}\frac{Q_{rot}}{J_uR_i}\frac{exp(E_u/kT_{ex})}{exp(h\nu/kT_{ex})-1}\int \tau_\nu dv
\label{column_density}
\end{eqnarray}
in which $\mu$ is the permanent electric dipole moment, Q$_{rot}\approx\frac{kT}{hB}+\frac{1}{3}$ denotes the partition function, R$_i$ is to the relative intensity of the transition of interest in relation to the total intensity, and E$_u$ is the energy of the upper state.

The above equation can be simplified assuming optically thin emission:
\begin{eqnarray}
N_{tot}(N_{2}H^{+})=\frac{6.25\times10^{15}}{\nu(GHz)R_i}\frac{exp(E_u/kT_{ex})}{exp(h\nu/kT_{ex})-1} \nonumber \\
\times \left[ \frac{T_B \Delta v(km~s^{-1})}{f(J_\nu(T_{ex})-J_\nu(T_{bg}))} \right] cm^{-2}
\end{eqnarray}
where $\nu$ is the frequency of the hyperfine transition used. In our case, the strongest hyperfine transition and main component F=(2,1), J=(1,0) is adopted, with $\mu$=3.37 Debye, $\nu$=93.1737767 GHz, B=46.5869 GHz, E$_u$=4.4716 K, $R_i=\frac{7}{27}$\footnote{The values are obtained from \href{http://www.cv.nrao.edu/php/splat/}{http://www.cv.nrao.edu/php/splat/}. Same for other spectrum.}, and $J_\nu(T)=\frac{h\nu/k}{exp(h\nu/kT)-1}$. T$_{ex}$ is adopted from the corresponding CO (1-0) observation when assuming LTE.

The values for $\tau_{main}$, N$_{N_{2}H^{+}}$, and $R_{N_{2}H^{+}}$ are presented in Table \ref{table:lines}. The range for N$_{N_{2}H^{+}}$ is 0.9--5.1$\times10^{13}$~cm$^{-2}$, which is akin to that observed in high-mass star-forming regions \citep{mie14}. The related $R_{N_{2}H^{+}}$ values vary over 0.7--2.6$\times10^{-9}$, which is comparable to the range of values obtained in massive clumps $1.2\times10^{-10}-9.8\times10^{-9}$ \citep{pir03,mar08,mie14}, regardless of whether the clumps are IR-bright or IR-dark.

\subsubsection{HCO$^+$ J=1-0}
\label{sec:hcop}
HCO$^+$ (1-0) emission was identified in each of our ten peaks, presented in black in Figure \ref{fig:spectrum}. The majority exhibit single peaks, while A2-6, A2-8 and A2-9 show distinct double peaks. The A2-8 and A2-9 present blue-asymmetric profiles, which are signatures of infall motion. If we assume these lines to be optically thick, they can be modeled to obtain properties such as the infall velocity v$_{in}$ and mass accretion rate $\dot{M}$ using the two-layer model of Myers et al. (1996). With a double Gaussian fit towards the HCO$^+$ line of A2-9, the peak intensities and central velocities of the red and blue components are 0.60~K at -0.37 km~s$^{-1}$, and 0.84~K at -1.52 km~s$^{-1}$, respectively. The corresponding intensity of the valley between two peaks is 0.34~K. The dip velocity, 1.20 km~s$^{-1}$, coincides with the velocity of the blue-shifted peak of $^{13}$CO (1-0), while there is no HCO$^+$ emission corresponding to the red-shifted CO component (see Section \ref{sec:2comps} for the analysis of cloud components). Following Equation 4 in \cite{tra18}, v$_{in}$ is derived as 0.10 km~s$^{-1}$. Assuming spherical geometry, the mass accretion rate is evaluated as 2.56 $\times$ 10$^{-5}$ M$_\odot$~yr$^{-1}$. This value is two orders of magnitude smaller than the average $\dot{M}$ of massive clumps (9.6 $\times$ 10$^{-3}$ M$_\odot$~yr$^{-1}$, \cite{tra18}), suggesting a relatively weak infall motion of our clump. Thus, the result may favor CCC rather than a global collapse of the S-shaped filament.

Alternatively, if we consider the HCO$^+$ line to be optically thin, the column density of a linear, rigid rotor molecule HCO$^+$ can be determined using Equation \ref{column_density}:
\begin{eqnarray}
N_{HCO^+} = \frac{3h}{8\pi^{3}\mu^{2}}(\frac{kT_{ex}}{hB}+\frac{1}{3})\frac{exp[hBJ(J+1)/kT_{ex}]}{J+1} \nonumber \\
\times \frac{exp(h\nu/kT_{ex})}{exp(h\nu/kT_{ex})-1}\frac{T_B \Delta v}{f(J_\nu(T_{ex})-J_\nu(T_{bg}))} 
\end{eqnarray}
where the value is $\mu$=3.888 Debye, $\nu$=89.188526 GHz, with B=44.5944 GHz, and E$_u$=4.28 K. In the absence of observations for transitions from an isotopologue or a higher excitation level, it is challenging to determine accurately the excitation temperatures of optically thin HCO$^+$ emission \citep{yua16}. Consequently, we assume it to be equivalent to $T_{ex}$ of CO (1-0), due to analogous excitation requirements \citep{fue05}. T$_B$ and $\Delta$v (FWHM) are derived from gaussian fitting by assuming a single component, listed in Table \ref{table:obs-fit}. The final HCO$^+$ column density is corrected for optical depth by multiplying the column density with a correction factor, $\tau$/(1--e$^{-\tau}$) \citep{gol99}.

Subsequently, the optical depth and column density are calculated, as illustrated in Table \ref{table:lines}. The column density of HCO$^+$ ranges over 1.4 to 3.6$\times10^{12}$~cm$^{-2}$, aligning with the established values for Planck clumps \citep{yua16}, albeit one order of magnitude lower than those observed in massive clumps \citep{san12,mie14}. Similar to N$_2$H$^+$, HCO$^+$ is predominantly more concentrated within the central cores compared to the cores located on either side of the filament. Furthermore, the abundance measured as 6.9$\times$10$^{-11}$--1.7$\times$10$^{-10}$ is marginally beneath the typical values for massive clumps, noted as $10^{-10}$--$10^{-8}$ \citep{san12,mie14}.

\subsubsection{H$_2$CO $2_{1,2}-1_{1,1}$}
The presence of H$_2$CO was identified in all the cores, characterized by peak intensities of 0.4--1.0~K (see red spectra in Figure \ref{fig:spectrum}, one-component gaussian fitting results in Table \ref{table:obs-fit}). Given its nature as a slightly asymmetric top molecule, nearly resembling a prolate top \citep{man15}, the column density of ortho-H$_2$CO in the 2(1,2)-1(1,1) transition can be estimated utilizing Equation \ref{column_density} with an alternative partition function Q$_{rot}=\sqrt{\frac{\pi (kT_{ex})^3}{h^3ABC}}$. The values for the rotational constants are A=281.97056 GHz, B=38.833987 GHz, C=34.004243 GHz, along with dipole moment $\mu$=2.33 Debye, $\nu$=140.83950 GHz, and E$_u$=21.92264 K. The final density is corrected with the optical depth as well. Furthermore, observations were confined to transitions of ortho-H$_2$CO, necessitating a correction of the column density through the implementation of an ortho/para ratio of 3:1, as suggested by \cite{rob02}.

As illustrated in Table \ref{table:lines}, all H$_2$CO lines are optically thin, allowing us to ascertain the column density. The value of N$_{H_2CO}$ ranges over 1.3--2.6 $\times$ 10$^{13}$~cm$^{-2}$, which is comparable to that reported in \cite{rob02} for excitation temperatures ranging from 10~K to 30~K. The presence of H$_2$CO suggests the existence of deeply embedded protostars within our cores \citep{fue05}. The range of $R_{H_2CO}$, specified as 6.4$\times10^{-10}$--1.5$\times10^{-9}$, is situated at the upper range of the abundance observed in Class 0 protostars (8$\times10^{-11}$--1$\times10^{-9}$ \cite{kan15}).

\section{discussion}
\label{sec:discussion}
\subsection{Observational evidence for CCC}
\label{sec:evidence}
With the interesting features observed of the G178 cloud, questions can be raised to the formation of the S-shaped filament, the mechanisms underlying the development of a series of cores within the filament, and the reasons behind the uniform spacing of these cores. This section examines various pieces of evidence supporting the occurrence of CCC in the G178 cloud.

\subsubsection{Two Components in CO spectrum}
\label{sec:2comps}
\begin{figure*}
    \centering
    \includegraphics[width=\linewidth]{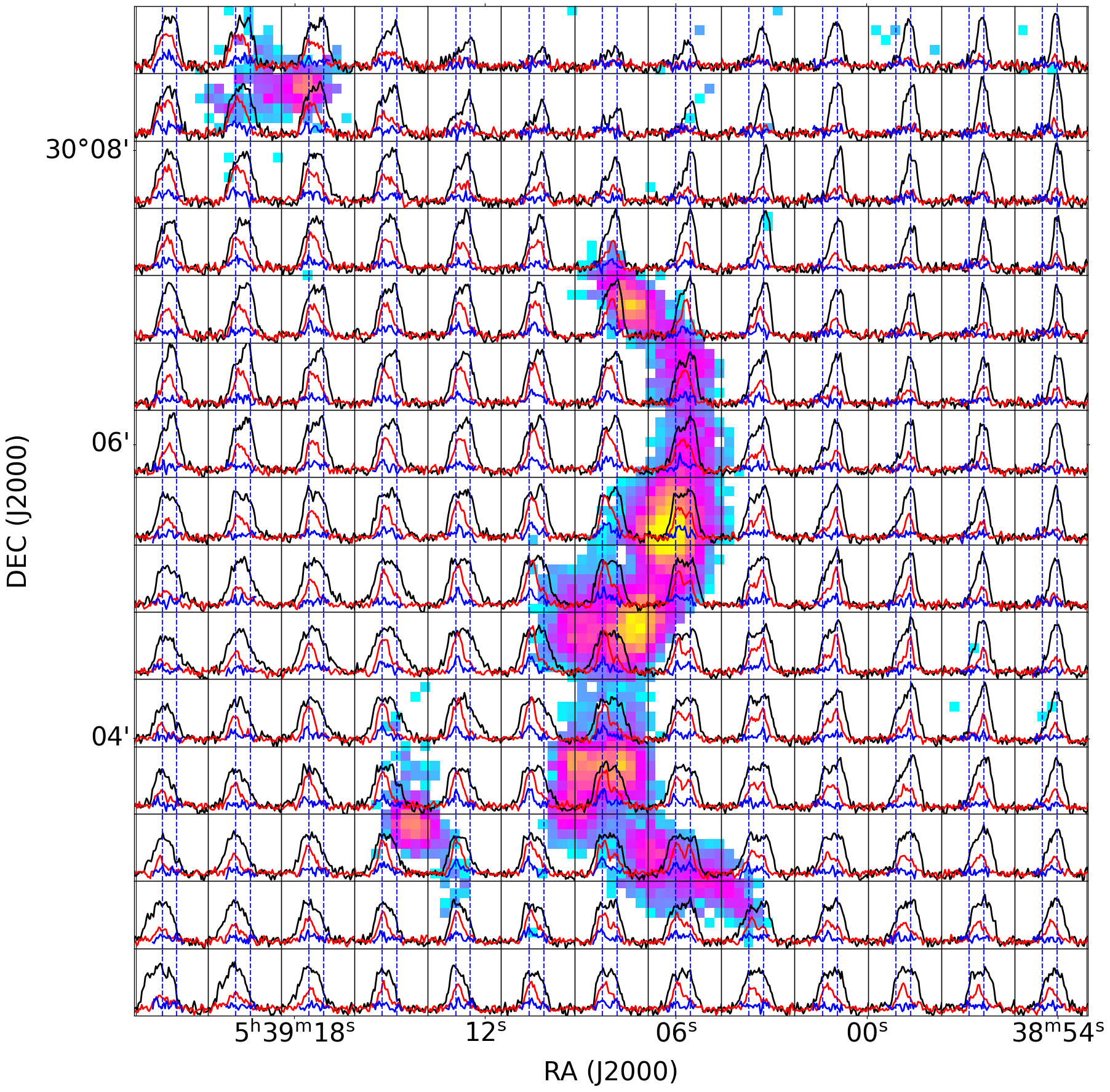}
    \caption{Pixel-by-pixel spectra of $^{12}$CO (1-0), $^{13}$CO (1-0) and C$^{18}$O (1-0) in black, red, and blue respectively. In each subplot, the x axis spans -5 km~s$^{-1}$ to 5 km~s$^{-1}$, and the y axis extends from -1~K to 12~K for $^{12}$CO (1-0), -1--8~K for $^{13}$CO (1-0) and -1--6~K for C$^{18}$O (1-0). The spectra exhibit two discernible velocity components, with their systemic velocities marked by blue dashed lines at -1.2 km~s$^{-1}$ and 0.8 km~s$^{-1}$. From east to west, $^{13}$CO (1-0) velocity structure transits from blue- to red-shifted components, with broad dual peaks at the central region that coincide with the 850 $\mu$m continuum shown as the background. The C$^{18}$O lines behave similar to the $^{13}$CO and peak at analogous velocities.}
    \label{fig:spec}
\end{figure*}

Distinct double peaks are observed in the lines of CO and its isotopologues for both J=1-0 and J=2-1 transitions, indicative of the presence of two distinct velocity components. Specifically, Figure \ref{fig:spec} shows the pixel-by-pixel spectra of $^{12}$CO (1-0) in black, $^{13}$CO (1-0) in red, and C$^{18}$O (1-0) in blue, superimposed on the 850 $\mu$m filament. In the field of view from the east to the west, the line profile of $^{13}$CO (1-0) transitions from a single peak at -1.2 km~s$^{-1}$ to a blue-asymmetric profile when approaching the filament, then to a symmetric doublet at the filament, and finally to a red-asymmetric profile before returning to a single peak at 0.8 km~s$^{-1}$. The systemic velocities of the two components, -1.2 km~s$^{-1}$ and 0.8 km~s$^{-1}$, are estimated from $^{13}$CO (1-0), one below and one above 0 km~s$^{-1}$ velocities, and separated by 2 km~s$^{-1}$, as indicated by the blue vertical line in each subplot of Figure \ref{fig:spec}. The observed spectra cannot be attributed to global infall, which would typically manifest as blue-asymmetric profiles across the cloud \citep{wu07}. Moreover, the systemic velocity of the blue-shifted $^{13}$CO (1-0) component is positioned between the dual peaks of HCO$^+$ at core A2-9 (-1.52 km~s$^{-1}$ and -0.37 km~s$^{-1}$, see Section \ref{sec:hcop}) and aligns with the dip velocity at -1.2 km~s$^{-1}$. This indicates that the blue-shifted CO component may represent a slowly collapsing clump, and that the double peaks observed in $^{13}$CO are unlikely to arise from the self-absorption of a single optically thick cloud (see the small $\tau$ of $^{13}$CO in Table \ref{table:CO}). The C$^{18}$O is characterized by a lower optical thickness compared to $^{13}$CO, yet it exhibits analogous behaviors. The observed C$^{18}$O peaks are always corresponding to the peak of $^{13}$CO at the same velocity, indicating these two lines are not heavily affected by optical depth. Consequently, the presence of two cloud components is supported in this analysis.

\begin{figure*}
    \centering
    \includegraphics[width=1\textwidth]{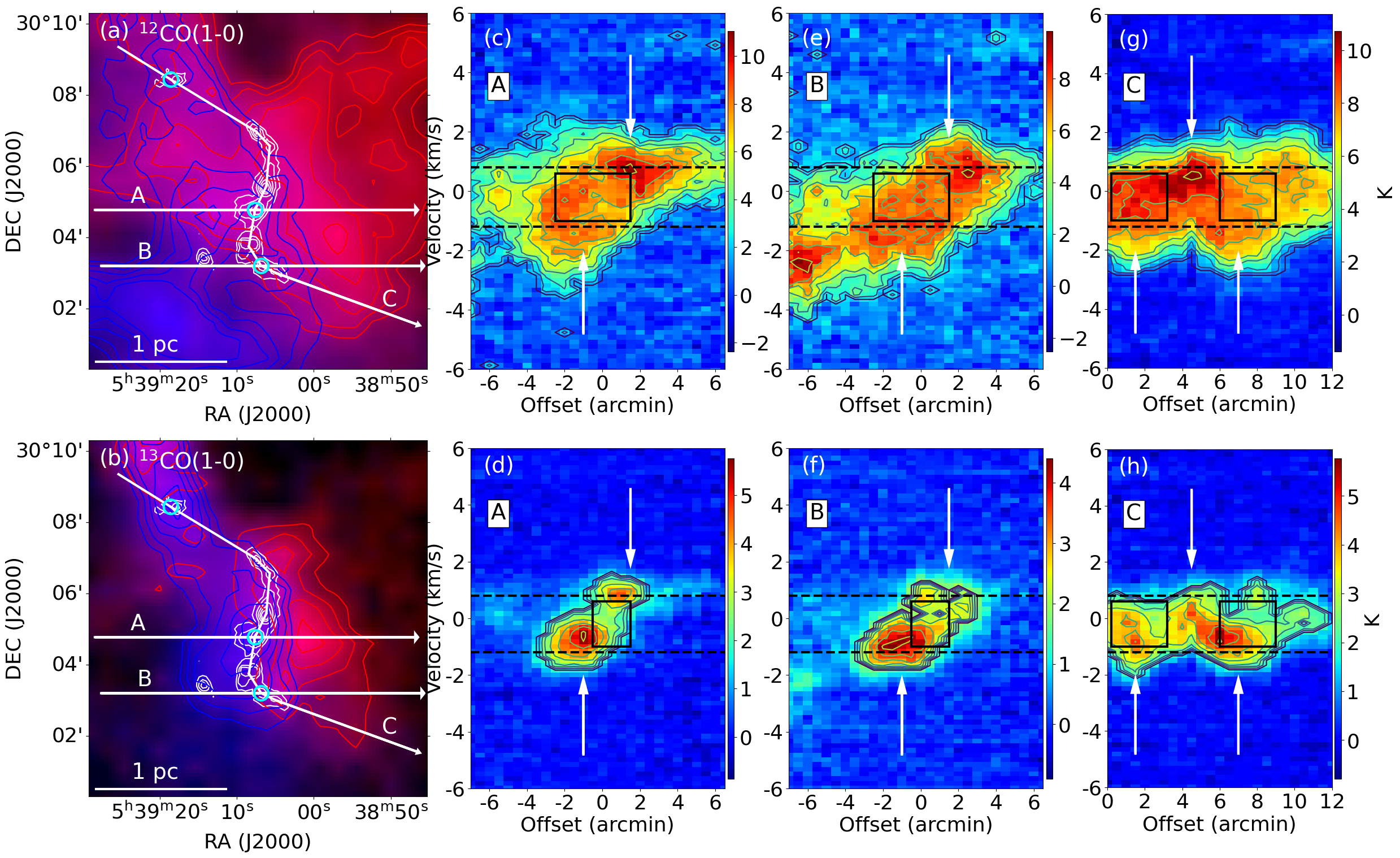}
    \caption{(a) A two-color composite integrated intensity map of $^{12}$CO (1-0), integrated from -5 km~s$^{-1}$ to 0 km~s$^{-1}$ (depicted in blue color and contours ranging over 16--24 K$\cdot$km~s$^{-1}$ with a step of 2~K$\cdot$km~s$^{-1}$) and from 0 km~s$^{-1}$ to 4 km~s$^{-1}$ (in red, contours ranging over 12--20 K$\cdot$km~s$^{-1}$ with a step of 2~K$\cdot$km~s$^{-1}$) of the G178 cloud. The cuts labeled A, B, and C identify the paths along which PV slices, with a width of one pixel, were extracted, sampling the core regions A2-5, A2-7, and the filament, respectively. (b) Similar as (a) but for $^{13}$CO (1-0). The contours for the blue-shifted cloud range over 8--16 K$\cdot$km~s$^{-1}$ with a step of 2~K$\cdot$km~s$^{-1}$, and for red-shifted cloud are 4--10 K$\cdot$km~s$^{-1}$ with a step of 2~K$\cdot$km~s$^{-1}$. Panel (c), (e), (g): PV diagrams of $^{12}$CO (1-0), along cuts A, B, and C. (d), (f), (h): PV diagrams of $^{13}$CO (1-0). The two horizontal dashed lines represents the systemic velocities of the blue-shifted cloud (-1.2 km~s$^{-1}$) and red-shifted cloud (0.8 km~s$^{-1}$). The bridging features are detected within the intermediate velocity range between the dashed lines, marked with black boxes. The peak positions of red and blue clouds are highlighted by white arrows on the PV cuts. Panels (g) and (h) show V-shaped structures noted by white arrows, indicating collisional interactions \citep{haw15}. The positions of possible collisional interactions are marked out by cyan circles in panels (a) and (b).}
    \label{fig:PV}
\end{figure*}

To explore the potential interaction between these two components, we examine the integrated intensity map of each component, illustrated in panels (a) and (b) of Figure \ref{fig:PV}. The blue and red lobes are integrated over the velocity range of -5 to 0 km~s$^{-1}$, and 0 to 4 km~s$^{-1}$, respectively. Notable features are identified in this visualization. The blue-shifted cloud dominates the eastern region, while the red-shifted cloud exhibits greater intensities towards the west. These clouds intersect in the center of the image, spanning an approximate linear scale of 1~pc (considering 3$\sigma$ overlap areas of the blue and red shifted components of $^{12}$CO (1-0)). Furthermore, the overlapping region encompasses the entire 850~$\mu$m continuum filament. Additionally, the ten continuum peaks are situated at the intersection of the two kinematic structures, rather than at their intensity maxima. 

The most plausible explanation for this phenomenon is a cloud-cloud collision. The separation between the two velocity components is measured at $\sim$2.0~km~s$^{-1}$, which is notably lower than the median velocity of 50 CCC clouds, recorded at 5~km~s$^{-1}$ (\cite{fuk21}). Instances of such low velocities have been documented in the context of CCC, as indicated by \cite{iss20}. According to \cite{fuk21}, a reduced collision velocity may facilitate the enhancement of low-mass star formation, a notion that is corroborated by our findings (refer to the mass discussion in Section \ref{sec:mass}). Assuming G178 exemplifies CCC, the collision timescale is estimated to be $\sim$0.5 Myr, derived from the division of 1.0~pc by 2.0 km~s$^{-1}$. This timescale is appropriate for star formation processes (for further details of associated star-forming activities, see Section \ref{sec:YSO}). Another possibility is the cloud rotation along the axis of vision, while it is not favored since the cloud is found to be not gravitationally bounded, according to the discussion of fragmentation conditions in Section \ref{sec:fragmentation}.

To determine the masses of the two clouds, we consider the region with $^{13}$CO (1--0) emission above 3$\sigma$ ($\sim$1.2~K) as the cloud complex, and assume that both clouds have equal contribution to the mass at the intersection. Subsequently, the total mass equals $(\Sigma N_{H_{2}}) \mu m_H A$, where $\Sigma N_{H_{2}}$ is the integral of the column density derived from SED within the cloud region, A is the physical size of the pixel, $\mu$ is the mean molecular weight, here 2.8, and m$_H$ is the mass of hydrogen atom. Thus, the individual masses of blue- and red-shifted clouds are estimated to be 210 M$_{\odot}$ and 160 M$_{\odot}$, respectively, with a total mass of 370 M$_{\odot}$. The small mass difference and low relative velocity support the formation of a gently-bent filament rather than a sharp arc-like interface, predicted by \cite{tak14} in Fig. 6 with v=3~km~s$^{-1}$ at maximum core-forming time.

A global-collapsing cloud may show double-peaked line features due to self-absorption, but it is not preferred in our case according to the following discussion. Considering an unmagnetized filament in virial equilibrium following Equation 12 in \cite{fie00}, the velocity dispersion necessary to maintain the equilibrium ($\sigma=\sqrt{G\cdot M/2L}$) for the observed filament (170~M$_\odot$, L$\sim$3~pc) is calculated as 0.8~km~s$^{-1}$. This value is less than the velocity separation of the double peaks, 2~km~s$^{-1}$. Consequently, this does not support the significant presence of supersonic turbulence or internal motions indicative of a collapsing cloud. Instead, it suggests a higher probability of observing a collision between two clouds.

\begin{figure*}
	\centering
	\includegraphics[width=\linewidth]{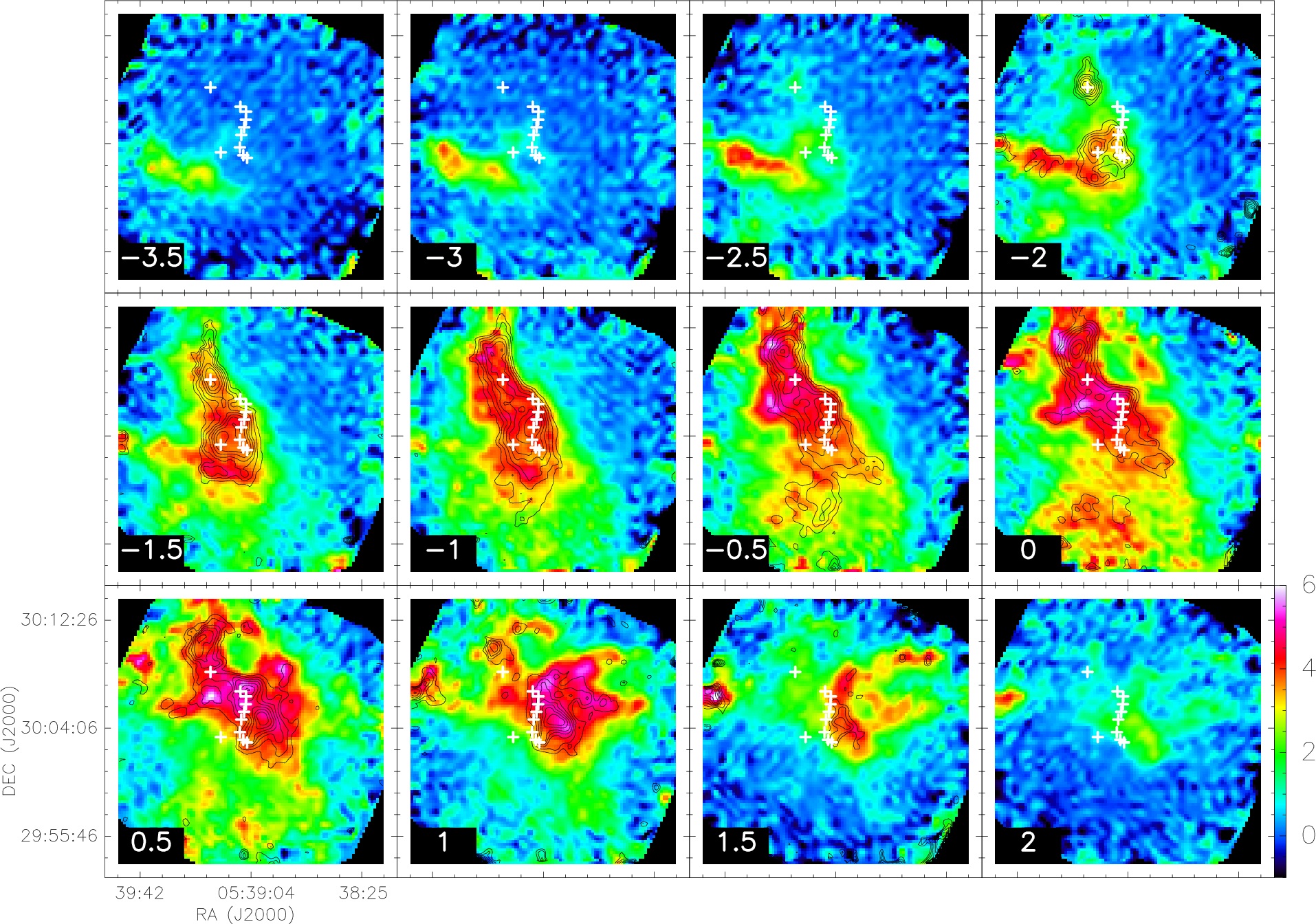}
	\caption{The channel maps of $^{12}$CO (1-0) (in color) and $^{13}$CO (1-0) (black contours) are shown. The color scale is in unit of K$\cdot$km~s$^{-1}$. Each map features a channel width of 0.5 km s$^{-1}$, with the corresponding velocities (in unit of km s$^{-1}$) marked at the bottom left corner of each subplot. The overlay of ten continuum peaks in white crosses effectively delineates the boundaries of the two colliding clouds.}
	\label{fig:channel_map}
\end{figure*}

\begin{figure*}
    \includegraphics[width=0.9\linewidth]{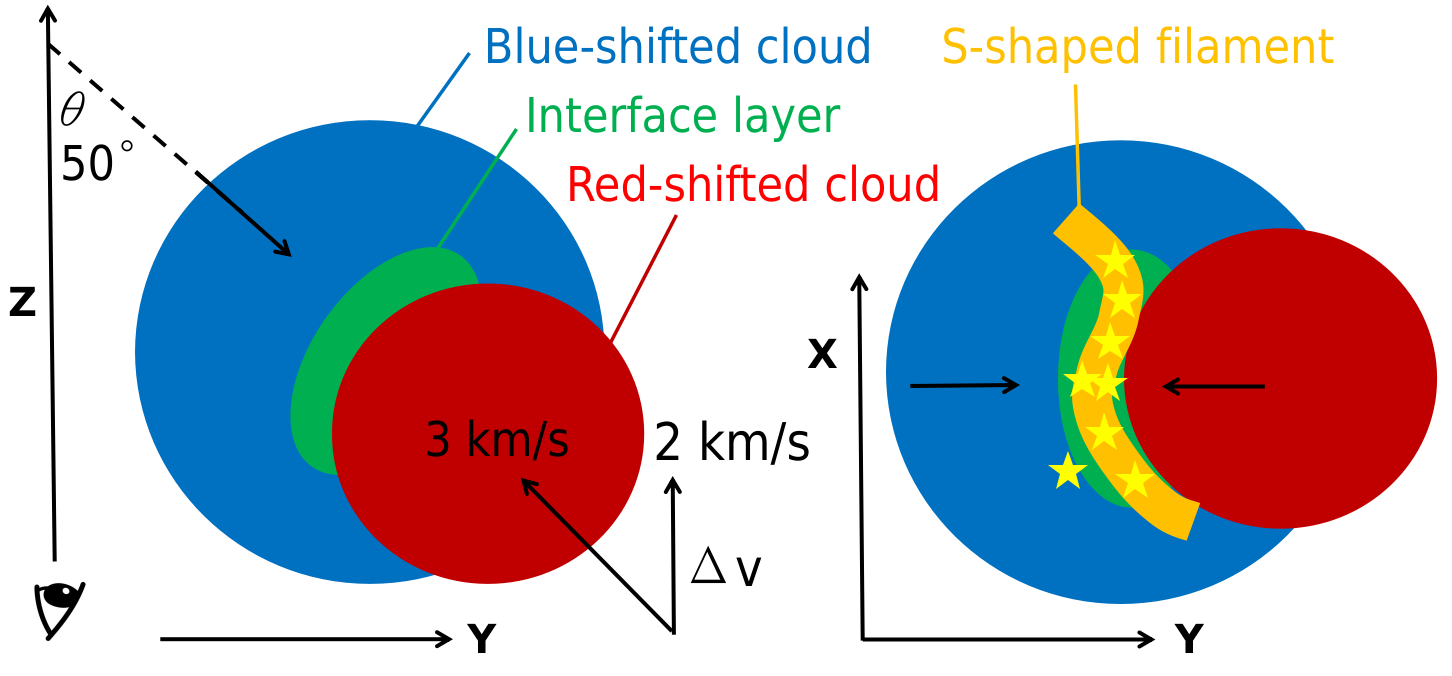}
    \centering
    \caption{Schematic illustrations of possible CCC in G178 as presented in the Z–Y plane (left panel) and in the X–Y plane (right), where the Z axis is along the line of sight. The collision axis is inclined at an angle of relative motion to the line of sight of $\theta$ = 50$^{\circ}$ (see Section \ref{sec:pv}).}
    \label{fig:sketch}
\end{figure*}

\begin{figure*}
	\centering
	\includegraphics[width=\linewidth]{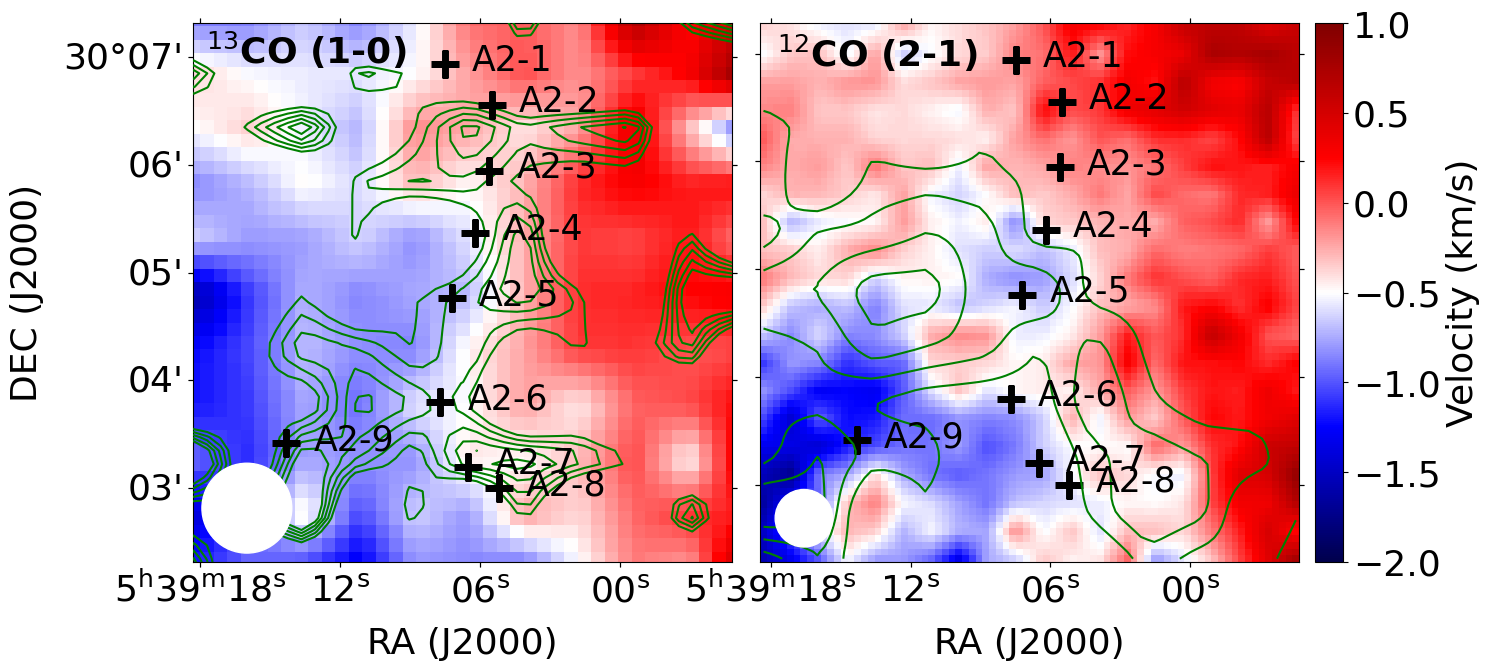}
	\caption{Moment 1 maps of $^{13}$CO (1-0) (left panel) and $^{12}$CO (2-1) (right panel) in color scale, with corresponding moment 2 maps overlaid as green contours. The contours for the line widths of $^{13}$CO (1-0) are 1.7--2.5 km~s$^{-1}$ with a step of 0.2 km~s$^{-1}$, and for $^{12}$CO (2-1) are 3.4--4.0 km~s$^{-1}$ with a step of 0.2 km~s$^{-1}$. The continuum peaks are labeled and found to locate at the intermediate velocity region. The beam sizes are shown as white circles.}
	\label{fig:moments}
\end{figure*}

\subsubsection{Channel maps}
Channel maps can elucidate intricate details of velocity components. Figure \ref{fig:channel_map} presents the channel maps of various CO lines at intervals of 0.5 km s$^{-1}$ bandwidth. In the CO (1-0) channel maps, the cloud emerges from the eastern side and progresses westward, reaching its maximum brightness in the central region. Each map is annotated with the locations of continuum cores, enabling the clear distinction of the edges of two clouds. Specifically, the continuum peaks precisely trace the boundaries of the red-shifted cloud as depicted in the 1.5 km s$^{-1}$ channel map and the blue-shifted cloud as shown in the -1.5 km s$^{-1}$ channel map. The red-shifted component is prevalent across four maps ranging from 0.5 km s$^{-1}$ to 2 km s$^{-1}$, whereas the blue-shifted component is apparent in several maps from -3.5 km s$^{-1}$ to 1 km s$^{-1}$. At 1.5 km s$^{-1}$, the smaller red-shifted cloud exhibits a U-shaped feature, with the bottom of the U shape aligning with the continuum filament. This complementary distribution with displacement signifies the presence of CCC, as predicted in simulations \citep{tak14, fuk21}. The U shape arises from the directed compression of a large cloud by a smaller one, concentrating the densest gas at the base of the U shape. Besides the main colliding components, the substructure elongated towards the northeast of the image may explain the formation of core A1. It may stretch the filament to depart from simple arc-like.

In the $^{13}$CO 1-0 channel maps that trace denser gas, the two velocity components exhibit a trend analogous to that observed in $^{12}$CO. The blue- and red-shifted clouds are spatially well separated, located to the ease and west of the S-shaped filament.

\subsubsection{PV diagrams}
\label{sec:pv}
Position-Velocity (PV) diagrams can further provide observational corroboration for CCC. Earlier simulations \citep{haw15} have demonstrated that bridge structures, which segregate two velocity peaks, are characteristic indicators of CCC in PV diagrams. In scenarios with lower collision velocities, however, these bridge structures are not easily discernible. In such cases, the PV diagram displays a V-shaped configuration, particularly for a collision model characterized by a velocity of 3 km~s$^{-1}$ at the time of maximum core formation, as illustrated in the lower panel of Fig. 5 in \cite{haw15}.

In this study, PV slices are extracted along three distinct directions (A, B, C) as illustrated in Figure \ref{fig:PV} (a) and (b). These slices serve to investigate cores A2-5 and A2-7, and the filament, respectively. Both $^{12}$CO (1-0) and $^{13}$CO (1-0) emission exhibit broad bridging features marked by black boxes in the (c) to (h) panels of Figure \ref{fig:PV}, which are located between the two systemic velocities, -1.2 km~s$^{-1}$ and 0.8 km~s$^{-1}$, delineated by black dashed lines. The blue- and red-shifted clouds are noted by the white arrows on the PV plots. The V-shaped structures, which are indicative of collisional interactions between two clouds of differing sizes \citep{haw15}, are discernible in Fig. \ref{fig:PV} (g) and (h), as indicated by white arrows, thereby delineating the locations of strong collision interface. These positions correspond to cores A1, A2-5, and A2-7, denoted by cyan circles in (a) and (b), thus providing evidence that the filament formation is a result of a CCC.

The velocity separation of 2 km~s$^{-1}$ provides a lower limit for the relative collision velocity. Due to the projection effects, however, the actual collision velocity typically surpasses this lower limit.
As suggested by \cite{iss20}, isotropic turbulence intensifies within the collision-shocked layer, regardless of the direction of the collision. As a result, the velocity dispersion within the shocked layer may be equated to the relative collision velocity. In accordance with this principle, we hypothesize that the relative velocity is equivalent to the FWHM of the $^{12}$CO (1-0) line observed at the peak position A2-5, designated as the representative collision site. The FWHM of this spectral line is determined to be 3.0 km~s$^{-1}$. By juxtaposing this measurement with the velocity difference between the two clouds, we deduce that the relative collision velocity translates to a relative motion of arccos(2.0/3.0) $\sim$50$^\circ$ with respect to the line of sight (see the sketch in Figure \ref{fig:sketch}).

\subsubsection{Velocity gradient}

Figure \ref{fig:moments} presents the moment one maps (vlsr) of $^{12}$CO (2-1) and $^{13}$CO (1-0). A notable gradient in the central velocity is observed between the two clouds, accompanied by a sharp delineation in the red section of the $^{13}$CO (1-0) map. The formation of most cores predominantly occurs in the counter part of red and blue components, which is presumably indicative of the area of collision. Additionally, the moment 2 maps (line width) are represented as green contours in Figure \ref{fig:moments}, displaying an enhanced line width along the filamentary structure, particularly for $^{13}$CO (1-0) that is optically thin, which implies the convergence of two clouds in the vicinity of the filament.

\subsubsection{Dust polarization}

The thermal dust polarization data for G178, including Stokes I, Q, and U maps at 353 GHz, were obtained exclusively from the Planck archive \citep{pla15}. From these data, we derive a polarization fraction of approximately $\sim$2\% $\left(\sqrt{Q^2+U^2}/I\right)$ and a linear polarization angle (PA) of around -12$^{\circ}$ ($\frac{1}{2}$arctan$\frac{U}{Q}$), measured clockwise from the Galactic north (or equal to 20$^{\circ}$ clockwise from the Equatorial west), as shown in panel (a) of Fig.~\ref{fig:continuum-SED}. Given that the angular resolution of the Planck observation is $\sim$5$'$ at 353 GHz, which is comparable to the entire size of the G178 filament, the above values represent an average measurement of the polarization across the cloud. Despite the limited spatial resolution, the derived polarization angle appears roughly perpendicular to the filament's overall orientation, particularly in its central part (core A2-4, A2-5). This configuration is generally consistent with the large-scale polarization morphology predicted by a CCC simulation (see Fig. 10 in \cite{kon24}). Furthermore, the simulation predicts polarization fractions between 1\% and 7\% in the filament's densest regions, which aligns well with the observed value of approximately 2\%.

While the current data provide encouraging evidence for the cloud–cloud collision scenario, existing observations remain insufficient to conclusively validate the model, particularly regarding the detailed mechanisms involved. High-resolution simulations by \cite{kon24} reveal significant small-scale variations in polarization angles ($\sim$10$''$) at the colliding interface, ranging from parallel to perpendicular relative to the filament. To enable a more direct comparison between observations and simulations, polarization measurements with comparable resolution, e.g., using JCMT/POL-2, are necessary. Moreover, current observational data and analysis methods are inadequate to determine whether the line-of-sight magnetic field reverses across the filament, as predicted by the model. Addressing these limitations in future work will be crucial for advancing our understanding of magnetic field configurations in filaments influenced by cloud–cloud collisions.

\subsection{Induced filament and star formation}

\subsubsection{Cylindrical fragmentation}
\label{sec:fragmentation}
The continuum cores are observed to be uniformly distributed along a ``cylinder'', though the filament is S-shaped. This pattern of fragmentation was initially accounted for by the phenomenon of ``sausage instability'' in a gas cylinder \citep{cha53}, and was elaborated upon by various subsequent studies \citep[e.g.,][]{wan14}. Under the assumption of an isothermal, hydrostatic, nonmagnetized, infinite gas cylinder maintained by both thermal and non-thermal pressure, it is feasible to compute the critical linear mass density M$_{cl}/\lambda_{cl}=465\left(\frac{\sigma}{1 km~s^{-1}}\right)^{2}M_\odot pc^{-1}$, the characteristic spacing between fragments $\lambda_{cl}=1.24pc\left(\frac{\sigma}{1 km~s^{-1}}\right)\left(\frac{n_c}{10^5 cm^{-3}}\right)^{-1/2}$, and the mass of fragments M$_{cl}=575.3M_\odot\left(\frac{\sigma}{1 km~s^{-1}}\right)^3\left(\frac{n_c}{10^5 cm^{-3}}\right)^{-1/2}$ \citep{wan14}. Here, $\sigma$ is adopted as the sound speed in the case of thermal support, and as the effective sound speed (1D velocity dispersion) in the case of non-thermal support; n$_c$ is the volume density at the center of the cylinder.

Considering the entire G178 filament as a cylinder, it is approximately 3~pc in length and 0.2~pc in width, with a total mass of 170 M$_{\odot}$ and a mean molecular weight of 2.8, leading to an average density of $2.3\times 10^{4}$~cm$^{-3}$ and a linear mass density of 57 M$_{\odot}$~pc$^{-1}$. If we only consider thermal pressure (assuming T = 13 K in $\sigma_{TH}=\sqrt{kT/\mu m_H}$), M$_{cl}/\lambda_{cl}$, $\sigma_{TH}$, M$_{cl}$, and $\lambda_{cl}$ are 0.2 km~s$^{-1}$, 19~M$_{\odot} pc^{-1}$, 10 M$_{\odot}$, and 0.5~pc, respectively. Although M$_{cl}$ is close to the observed value of G178 (core mass $\sim$ 9.7$\pm$7.5~M$_{\odot}$), the predicted core spacing $\lambda_{cl}$ is approximately 3 times larger than the observed mean core spacing (0.17~pc), indicating that thermal pressure alone may not govern the fragmentation process. When turbulent support is also taken into account, the effective sound speed $\sigma_{eff}$ can be estimated from the average C$^{18}$O line width (FWHM$\sim$1.3~km~s$^{-1}$) using the following relations: $\sigma_{C^{18}O}=\frac{FWHM}{\sqrt{8ln2}}$, $\sigma_{NT}=\sqrt{\sigma^2_{C^{18}O}-kT/m_{C^{18}O}}\sim$0.5~km~s$^{-1}$, $\sigma_{eff}=\sqrt{\sigma^2_{TH}+\sigma^2_{NT}}$. Taking $\sigma_{eff}\sim$0.6~km~s$^{-1}$, the corresponding values for 
M$_{cl}/\lambda_{cl}$, M$_{cl}$, and $\lambda_{cl}$ become 160~M$_{\odot}$~pc$^{-1}$, 240 M$_{\odot}$, and 1.5~pc, respectively.
The inclusion of turbulence increases the value of $\lambda_{cl}$ even further. Additionally, the critical mass per unit length exceeds the observed filament mass per unit length (57~$M_{\odot}$~pc$^{-1}$) by a factor of approximately three, suggesting that the G178 cloud is gravitationally unbound and, under cylindrical fragmentation theory, should not fragment into individual cores. Therefore, the observed fragmentation in this cloud cannot be explained by this mechanism alone. Additional physical processes must contribute to binding the system and driving core formation. A plausible explanation is the CCC scenario, which is supported by the observational evidence discussed in Section~\ref{sec:evidence}.

According to the magnetohydrodynamics simulations of CCC \cite{kon24}, a prediction is made regarding the gravitationally unbounded filament under the CMR mechanism. In this scenario, shocked gas undergoes condensation and is conveyed to the main filament incrementally through the action of reconnected magnetic fields, rendering the process of filament and star formation as a bottom-up process. This hypothesis may explain the high level of fragmentation under intense turbulence within the G178 filament. Furthermore, \citep{kon24} predicted a 2~pc S-shaped filament, depicted in their Figure 9(g) for a specific viewing angle, which is analogous in both morphology and physical scale (3 pc) to that of G178. 
Moreover, their simulations of $^{12}$CO with the use of RADMC-3D (in section 3.4.3) demonstrate that CO is marginally optically thick ($\tau\sim$2) at a column density of 10$^{17}$~cm$^{-2}$. In the dense filament, the CO column density reaches up to 3.3$\times$10$^{20}$~cm$^{-2}$, indicating optically thick conditions, while regions farther from the filament exhibit column densities below 10$^{17}$~cm$^{-2}$ and remain optically thin. Our observations (see Table \ref{table:CO}) yield a CO column density of (8.25$\pm$1.07)$\times$10$^{17}$~cm$^{-2}$, consistent with the prediction of \citep{kon24}. The derived optical depths, $\tau\sim$40, representing the densest region of the filament, are broadly consistent with the simulation.

We also compare the observations with other CCC models to assess their relevance to G178. \citet{tak14} simulated the formation of a C-shaped filament resulting from the interaction between a large cloud and a smaller one at a relative velocity of 3~km~s$^{-1}$. While this model captures some filamentary features, it does not account for the S-shaped morphology observed in G178. Model 4 of \citet{wu17} reproduces a more similar filament structure; however, it assumes a non-head-on collision between two identical clouds with a much higher relative velocity of 20~km~s$^{-1}$, a scenario inconsistent with the physical conditions inferred for G178. In contrast, the model proposed by \citet{kon24} offers the closest match to our observations, both in morphology and kinematics, although the magnetic fields remain unconstrained in observation.

\subsubsection{Evolutionary stage}
\label{sec:evolution}
As evidenced by Table \ref{table:lines}, it is observed that N$_{N_{2}H^{+}}$ and $R_{N_{2}H^{+}}$ values tend to be increased within the denser central region of the filament. This phenomenon may be attributed to the large beam size of KVN (37$''\sim$~0.17~pc), which captures the cold N$_2$H$^+$ gas surrounding the comparatively warmer central YSOs \citep{san12}. Alternatively, it may result from the differential freeze out of CO, i.e., CO depletion. According to the following reaction:
\begin{equation}
\label{equ:depletion}
N_{2}H^{+} + CO \rightarrow HCO^+ + N_2,
\end{equation}
N$_2$H$^+$ will be more abundant if CO is frozen on ice. Such phenomenon is seen in Fig. \ref{fig:moments}, as core peaks A2-6 and A2-9 show possible CO depletion since they do not correlate with emission peaks of C$^{18}$O (2-1) moment 0 map, leading to relatively high density and abundance ratio of N$_2$H$^+$ than the other cores.
In light of findings by \cite{hoq13}, which indicate that $R_{N_{2}H^{+}}$ increases with the evolutionary stage, it is highly probable that the central cores in our study are at relatively advanced evolutionary stages compared to the cores located on either end of the filament.

The ratio of N$_2$H$^+$ to HCO$^+$ column density serves as an indicator of the chemical evolutionary phase \citep{liu18,san12}. During the initial stages, N$_2$H$^+$/HCO$^+$ exhibits elevated levels due to the freezeout of CO and the subsequent depletion of HCO$^+$. In a warmer environment, specifically when the temperature approaches T$\sim$20~K, CO molecules are released from dust grains, consequently leading to the potential destruction of N$_2$H$^+$, while the abundance of HCO$^+$ is augmented via the reaction \ref{equ:depletion}.
Consequently, a lower abundance ratio of N$_2$H$^+$/HCO$^+$ could suggest that the clumps have reached a even more advanced stage of chemical evolution. \cite{aik05} predict that abundance ratios of N$_2$H$^+$/HCO$^+$ will fall within the range of 0.02$\sim$0.1 under the physical conditions pertinent to pre-stellar collapsing dense cores. This chemical model aligns with observations noted in starless cores such as L1544 \citep{cas02} and in regions of high mass star formation \citep{hoq13}.

Nevertheless, the N$_2$H$^+$/HCO$^+$ ratio observed here appears to exceed expectations by 2 to 3 orders of magnitude, as illustrated in the last column of Table \ref{table:lines}, displaying a range from 3.8 to 12.1. This discrepancy might arise from an overestimation of N$_2$H$^+$ column density, whereas HCO$^+$ might be underestimated when utilizing T$_{ex}$ from CO, or HCO$^+$ line tends to be optically thick. 
Therefore, evolution in our case might not be well traced by the N$_2$H$^+$/HCO$^+$ ratio. This conclusion, which is contrary to the theoretical expectation, is also reported in \cite{hoq13}, where a slight increase in the N$_2$H$^+$/HCO$^+$ abundance ratio was seen as evolution progresses in high-mass star-forming clouds. 

\subsubsection{Induced star formation}
CCC-triggered star formation is a plausible scenario in G178. We discuss how CCC facilitates the fragmentation and subsequent (high-mass) star formation based on the following evidence.

Given that G178 is an elongated filament exhibiting sub-Jeans fragmentation (or a stable condition), it is possible to ascertain the timescales associated with its fragmentation and collapse. According to the Equation 13 in \cite{hac25}, the fragmentation timescale of G178 is determined to be approximately 0.34 Myr, based on an average density of 2.34$\times10^{4}$ cm$^{-3}$ along the filament. This is marginally less than the colliding timescale of 0.5 Myr, suggesting that the fragmentation could potentially be a consequence of collision. Furthermore, the timescale for the longitudinal collapse, defined as the duration required for collapse along the principal axis of the filament, can be formulated as a function of the free-fall time, as expressed in Equation 16 in \cite{hac25}. Given the aspect ratio of 15 (length of 3 pc and width of 0.2 pc) for G178, the longitudinal collapse time is estimated to be approximately 1.5 Myr, shorter than the average lifespan of Class II YSOs \citep[$\sim$2 Myr,][]{eva09}. Despite this, Class II YSOs have indeed been detected, with the most prominent example being associated with the distinctive core A2-5 at the filament's center. The formation of a cluster of Class II YSOs is characteristically observed at the center of a hub-filamentary system \citep{vaz19}. However, this scenario is excluded for the G178 filament due to the lack of an evident hub structure and the absence of a discernible velocity gradient along the filament (refer to moment 1 maps in Fig. \ref{fig:moments}). Consequently, the rapid evolution of the cores into YSOs, as opposed to gravitationally dominated scenarios, may indicate star formation induced by CCC. In addition, the estimated colliding timescale of 0.5 Myr aligns closely with the typical age of Class I YSOs \citep[0.6~Myr,][]{eva09}. This similarity provides compelling evidence that the youngest stellar populations may have been initiated by a collapse approximately 0.5 Myr ago, further corroborated by the absence of detected YSOs in the regions where the two clouds do not intersect. 

Secondly, G178 is anticipated to generate a cluster comprising low- to intermediate-mass stars. Despite the detected dust masses (2.2 to 21.3 M$_\odot$) and a star formation efficiency between 6\% and 9\% \citep{eva09}, it may be inadequate to facilitate the formation of intermediate-mass stars. Nevertheless, there remains a potential for the formation of intermediate-mass stars. The column densities of all cores situated in G178 exceed 1 $\times$ 10$^{22}$ cm$^{-2}$ at a scale of 0.2 pc, suggesting the plausibility of high-mass star formation as indicated by \cite{fuk21}. Moreover, in four out of the nine cores, the surface densities surpass 0.1 g~cm$^{-2}$, further substantiating the indication of high-mass stellar formation \citep{tra20}. The presence of large Jeans masses also necessitates the further accumulation of mass, thereby creating an optimal condition for the formation of higher mass stars.

\section{Summary}
\label{sec:summary}
The Planck Cold Clump, G178.28-00.61, has been observed at multiple wavelengths. The continuum observations include 450 $\mu$m and 850 $\mu$m images captured by the JCMT, along with infrared archival data from WISE and Herschel. For molecular lines, it was mapped in $^{12}$CO,$^{13}$CO,C$^{18}$O J=1-0 lines by PMO and J=2-1 by CSO. Futhermore, single-point spectral line observations of H$_2$O (6-5), CH$_3$OH (7-6), HCO$^+$ (1-0), N$_2$H$^+$ (1-0), and H$_2$CO ($2_{1,2}-1_{1,1}$) at ten 850 $\mu$m continuum peaks were conducted by KVN, though H$_2$O and CH$_3$OH was below the detection limit.

The investigation identified two CO molecular components moving at distinct velocities (-1.2 km s$^{-1}$ and 0.8 km s$^{-1}$) coexisting with an 850 $\mu$m S-shaped continuum filament in a 1.0~pc-wide region of overlap. Ten continuum peaks along the filament exhibit a regular spacing of 0.17$\pm$0.01~pc, which cannot be explained by turbulence-induced cylindrical fragmentation or Jeans fragmentation. The dust temperatures (12.4$\pm$0.2~K) and column densities (1.9$\pm$0.5$\times 10^{22}$~cm$^{-2}$) for the nine cores have been calculated using two-component SED fitting, resulting in mean core masses of 9.7$\pm$7.5~M$_{\odot}$. Among the cores, eight are associated with Class I YSOs and one with a Class II YSO. The inner cores appear to be more evolved relative to those at the ends of the filament, as indicated by the column density and the abundance of dense molecular tracers.

We postulate that the two clouds have undergone a collision in this region, driven by the following observed phenomena: 
(1) the transition in the CO velocity structure from a blue-shifted to a red-shifted component, accompanied by dual peaks at the central region that coincide with the 850 $\mu$m S-shaped filament,
(2) spatially separated blue- and red-shifted clouds, located to the east and west of the filament, as illustrated in the channel maps,
(3) the bridge structures presented in the position-velocity diagrams, correlating with the shock-compressed layer anticipated by cloud-cloud collision simulations, 
(4) the elevated fragmentation level exceeding that expected from thermal and turbulent support, and 
(5) the enhanced star formation activity at the center of the filament.

Figure \ref{fig:sketch} illustrates the proposed collision scenario in G178, where a smaller red cloud (160~$M_{\odot}$) collides with a larger blue cloud (210~$M_{\odot}$) at a relatively mild collision velocity of approximately 3.0~km~s$^{-1}$ (corrected for a viewing angle of 50$^{\circ}$). This interaction leads to the formation of a compressed layer and an S-shaped filament, subsequently triggering star formation. Given these observational characteristics, G178 represents a valuable target for studying the early stages of cloud–cloud collision prior to significant influence from stellar feedback. The dynamical impact of the collision is likely to persist in the future evolution of the entire system.

The simulation presented in \citet{kon24} may substantiate the conclusion regarding CCC in G178 by highlighting their analogous characteristics in terms of morphology, physical scale, and density, as well as the dust polarization perpendicular to the filament, thus permitting the G178 filament to remain gravitationally unbound through the mechanism of collision-induced magnetic reconnection. With further high-resolution observations of molecular lines, dust continuum, and polarization, various simulations can be investigated to elucidate the observed phenomena.

\section*{Acknowledgment}
T. Z.\ gratefully acknowledges support by the National Natural Science Foundation of China (grant No. 12373026), the Leading Innovation and Entrepreneurship Team of Zhejiang Province of China (grant No. 2023R01008), the Key R\&D Program of Zhejiang, China (grant No. 2024SSYS0012), the China Postdoctoral Science Foundation No. 2023TQ0330, and thanks Jinghua Yuan and Sungju Kang for their constructive ideas and technical help.
T. L.\ gratefully acknowledges the supports by the National Key R\&D Program of China (No. 2022YFA1603101), National Natural Science Foundation of China (NSFC) through grants No.12073061 and No.12122307, the PIFI program of Chinese Academy of Sciences through grant No. 2025PG0009, and the Tianchi Talent Program of Xinjiang Uygur Autonomous Region.
L.B.\ gratefully acknowledges support by the ANID BASAL project FB210003. 
D.J.\ is supported by NRC Canada and by an NSERC Discovery Grant.
RKY gratefully acknowledges the support from the Fundamental Fund of Thailand Science Research and Innovation (TSRI) (Confirmation No. FFB680072/0269) through the National Astronomical Research Institute of Thailand (Public Organization). X.L. gratefully acknowledges support from the Strategic Priority Research Program of the Chinese Academy of Sciences under Grant No. XDB0800303.
We are grateful to the staff at the Qinghai Station of PMO, JCMT, CSO, and KVN for their assistance during the observations. The JCMT is operated by the East Asian Observatory on behalf of the National Astronomical Observatory of Japan, Academia Sinica Institute of Astronomy and Astrophysics, the Korea Astronomy and Space Science Institute and Center for Astronomical Mega-Science (as well as the National Key Research and Development Program of China with No. 2017YFA0402700). This research was carried out in part at the Jet Propulsion Laboratory, which is operated by the California Institute of Technology under contract with the National Aeronautics and Space Administration. This publication has used data products from the Wide-field Infrared Survey Explorer, which is a joint project of the University of California, Los Angeles, and the Jet Propulsion Laboratory/California Institute of Technology, funded by the National Aeronautics and Space Administration.

%\clearpage

\clearpage
\end{document}